\documentclass[aps,prb,twocolumn,amsmath,amssymb,nofootinbib,eqsecnum,tighten,showpacs,floatfix]{revtex4}

\usepackage[dvips]{color} 
\usepackage{graphicx}
\usepackage{dcolumn} 
\usepackage{bm}      
\usepackage{epic}
\usepackage{longtable}
\usepackage{curves}

\begin{document}

\title{
Influence of higher \mbox{$d$-wave} gap harmonics on the dynamical 
magnetic susceptibility of high-temperature superconductors
      }

\author{A.\ P.\ Schnyder}
\affiliation{Paul Scherrer Institute,
             CH-5232 Villigen PSI, 
             Switzerland}
\author{A.\ Bill}
\affiliation{Paul Scherrer Institute,
             CH-5232 Villigen PSI, 
             Switzerland}
\author{C.\ Mudry}
\affiliation{Paul Scherrer Institute,
             CH-5232 Villigen PSI, 
             Switzerland}
\author{R.\ Gilardi}
\affiliation{Laboratory for Neutron Scattering,
             ETH Zurich and PSI Villigen,
             CH-5232 Villigen PSI, 
             Switzerland}
\author{H. M. R\o nnow}
\affiliation{Laboratory for Neutron Scattering,
             ETH Zurich and PSI Villigen,
             CH-5232 Villigen PSI, 
             Switzerland}
\author{J.\ Mesot}
\affiliation{Laboratory for Neutron Scattering,
             ETH Zurich and PSI Villigen,
             CH-5232 Villigen PSI, 
             Switzerland}

\date{\today}

\begin{abstract}
Using a fermiology approach to the computation of the magnetic susceptibility
measured by neutron scattering in hole-doped high-$T^{\ }_c$ superconductors,
we estimate the effects on the incommensurate peaks
caused by higher \mbox{$d$-wave} harmonics of
the superconducting order parameter induced by underdoping.
The input parameters for the Fermi surface and \mbox{$d$-wave} gap are taken
directly from angle resolved photoemission (ARPES) experiments
on Bi$_2$Sr$_2$CaCu$_2$O$_{8+x}$ (Bi2212).
We find that higher \mbox{$d$-wave} harmonics lower the
momentum dependent spin gap at the incommensurate peaks
as measured by the lowest spectral edge
of the imaginary part in the frequency dependence 
of the magnetic susceptibility of Bi2212. 
This effect is robust whenever the fermiology approach 
captures the physics of high-$T^{\ }_c$ superconductors.
At energies above the resonance we observe diagonal incommensurate peaks.
We show that the crossover from parallel incommensuration below the resonance 
energy to diagonal incommensuration above it
is connected to the values and the 
degeneracies of the minima of the \mbox{2-particle} energy continuum.
 \end{abstract}

\maketitle

\section{
Introduction
        }
\label{Introduction}

The imaginary part 
$\chi^{\prime\prime}(\omega, \mathbf{q})$ 
of the magnetic susceptibility 
$\chi(\omega, \mathbf{q})$ 
probed by neutron scattering
in high-temperature (high-$T^{\ }_c$) 
superconductors is characterized by a very
rich dependence on energy and momentum transfer, temperature, and doping.%
\cite{rossat91a,mook93,sternlieb94,fong95,dai96,bourges96,fong97,dai98,%
      mook98,dai99,arai99,bourges00,fong00,mook00,dai01,pailhes03,stock04,%
      pailhes04,hayden04,
      shirane89,cheong91,mason93,yamada95,hayden96,mason96,aeppli97,%
      yamada98,
      lake99,lee00,gilardi04,christensen04,
      fong99,mook98bis,mesot00,he01,he02}
At temperatures well below the superconducting transition temperature 
$T^{\ }_c$ and for a fixed frequency $\omega$ 
(of the order of the \mbox{$d$-wave} gap maximum around optimal doping)
within a finite range of frequencies,
$\chi^{\prime\prime}$ 
displays peaks at some symmetry related wave vectors 
in the Brillouin zone of the 
square lattice formed by planar Cu sites.
The positions, heights, and widths of these peaks are temperature dependent
and, in particular, are sensitive to the destruction
of superconductivity above $T^{\ }_c$ 
to a degree that depends on doping.
As the energy transfered from the neutrons to the sample
is varied, 
the positions in the Brillouin zone of the peaks in 
$\chi^{\prime\prime}$ 
also vary (as well as their heights and widths), i.e., 
the peaks are dispersing. 
The detailed intensity distribution of the dynamical
spin susceptibility depends on the 
high-$T^{\ }_{\mathrm{c}}$ superconducting family.

For the
YBa$_2$Cu$_3$O$_{6+x}$ (YBCO) compounds
it is observed that the
separation in~\mbox{$\mathbf{q}$ space} between four
 incommensurate peaks at the symmetry related
wave vectors~$\mathbf{q}=(\pi \pm \delta, \pi)$ 
and~$\mathbf{q}=(\pi, \pi \pm \delta)$ decreases 
with increasing energy merging into a single 
resonance peak
at the antiferromagnetic wave vector $(\pi,\pi)$
and at an energy of~$41meV$ for optimally doped samples.%
\cite{rossat91a,mook93,sternlieb94,fong95,dai96,bourges96,fong97,%
      dai98,mook98,dai99,arai99,bourges00,fong00,mook00,dai01,%
      pailhes03,stock04,pailhes04,hayden04}
Cooling below the superconducting transition temperature opens up
a doping-dependent spin gap,
which is proportional to~$T^{\ }_c$.\cite{dai01}

In the La$_{2-x}$Sr$_x$CuO$_4$ (LSCO) family the dynamical
magnetic susceptibility follows a similar dispersion.
At energies above a spin gap  four 
incommensurate peaks on the horizontal and vertical lines
passing through $(\pi, \pi)$ in the Brillouin zone appear
and disperse 
towards the antiferromagnetic zone center as the
energy transfer is increased.
\cite{shirane89,cheong91,mason93,yamada95,hayden96,mason96,aeppli97,yamada98,%
      lake99,lee00,gilardi04,christensen04}
Although the incommensurate peaks in LSCO
seem to join at the commensurate 
wave vector~$\mathbf{q}=(\pi, \pi)$, no resonance peak has been observed 
in this compound to this date. Another remarkable observation
in LSCO is that a spin gap has not been observed 
in the extreme  underdoped and overdoped regimes.\cite{lee00,gilardi04}
 
On 
Bi$_2$Sr$_2$CaCu$_2$O$_{8+x}$ (Bi2212)
and Tl$_2$Ba$_2$CuO$_{6+x}$ (TBCO) samples only a few inelastic
neutron scattering measurements have been performed so far. For both compounds 
a resonance peak was observed,
\cite{fong99,mook98bis,mesot00,he01,he02}
but an incommensurate signal below the resonance has not yet
been measured, perhaps due to the limited size of single-crystals presently 
available.

In this paper we will take the point of view that incommensurate 
and commensurate peaks of 
$\chi^{\prime\prime}$  
have a common origin as they appear to be smoothly connected in the
YBCO family.
\cite{mook98,arai99,bourges00,dai01,batista01}
This hypothesis seems hard to reconciliate with
theoretical scenarios based on the existence of dynamical stripes
or on the SO(5) approach to high-$T^{\ }_c$ superconductivity.
In the dynamical stripe scenario incommensurate
peaks are the natural descendants of the static charge and 
spin long-range order seen in
La$_{2-x}$Ba$_x$CuO$_4$ at $x=1/8$,%
\cite{tranquada04} 
say.%
\cite{machida89,poilblanc89,zaanen89,schulz89,emery93,kivelson03,uhrig04}
In the SO(5) scenario a resonance at $(\pi,\pi)$ 
appears naturally as a result of a bound state in the spin-triplet
particle-particle channel.%
\cite{demler95,zhang97,greiter97,demler98,brinckmann98,tschernyshyov01,jiang-ping01}
Commensurate and incommensurate peaks
of
$\chi^{\prime\prime}$ 
are smoothly connected in a scenario in which it is assumed that
strongly renormalized quasiparticles close to the Fermi surface interact with
a residual on-site Hubbard or nearest-neighbor antiferromagnetic 
interaction, in short a fermiology approach. However it remains to be shown
if any disagreement between the line shapes for the magnetic susceptibility
predicted by fermiology and the measured line shapes is a mere 
quantitative one that results from the inherent
approximations involved in fermiology or is of a more qualitative nature
that reflects physics beyond the reach of fermiology.

Common to all fermiology scenarios 
\onlinecite{bulut90}--\onlinecite{ito04}
(see also Refs.\ \onlinecite{lu92,kee99,voo00} for related works
on the ``bare'' magnetic susceptibility)
is the random phase approximation (RPA) 
\begin{eqnarray}
\chi^{\vphantom{\prime}}_{\mathrm{RPA}}(\omega,\mathbf{q})=
\frac{
g(\omega,\mathbf{q})
     }
     {
1
+
h(\omega,\mathbf{q})
\Pi(\omega,\mathbf{q})
     }
\label{eq: RPA chi in fermiology}
\end{eqnarray}
to the magnetic susceptibility.
Here, 
$\Pi(\omega,\mathbf{q})$ is the (``bare'') magnetic susceptibility of
noninteracting fermionic (BCS) quasiparticles that depends sensitively on 
the Fermi surface (the superconducting BCS gap) above (below) $T^{\ }_c$.
The functions $g(\omega,\mathbf{q})$ and $h(\omega,\mathbf{q})$
are model dependent: A RPA treatment of the 
single-band Hubbard model with on-site repulsion $U$ yields
$g(\omega,\mathbf{q})=\Pi(\omega,\mathbf{q})$
and 
$h(\omega,\mathbf{q})=-U$.%
\cite{bulut90,schulz90,bulut92,bulut93,benard93,lavagna94,marel95,bulut96,salkola98,norman00,norman01}
A RPA treatment of a single-band of fermionic (BCS) quasiparticles
with a residual interaction such as
a nearest-neighbor antiferromagnetic interaction of strength
proportional to $J$ yields
$g(\omega,\mathbf{q})=\Pi(\omega,\mathbf{q})$
and 
$h(\omega,\mathbf{q})=(J/2)(\cos q^{\ }_x+\cos q^{\ }_y)$,%
\cite{maki94,mazin95,blumberg95,li98}
as is often done in the slave-boson treatment%
\cite{tanamoto91,tanamoto93,tanamoto94,stemmann94,brinckmann99,yamase00,li00,yamase01,brinckmann01,yamase02,li02,yamase03,li03}
or the $1/z$ expansion 
with $z$ the number of nearest-neighbors of the $t-J$ model.%
\cite{onufrieva95,onufrieva99,onufrieva00,onufrieva02}
A RPA treatment on the bare propagator 
$\chi^{\ }_0(\omega,\mathbf{q})$
of collective spin-1 excitations 
due to a weak coupling as measured by the coupling constant
$\mathrm{g}$
with otherwise noninteracting fermionic (BCS) quasiparticles
yields
$g(\omega,\mathbf{q})=\chi^{\ }_0(\omega,\mathbf{q})$
and 
$h(\omega,\mathbf{q})=-\mathrm{g}^2\chi^{\ }_0(\omega,\mathbf{q})$.%
\cite{morr98,abanov99,morr00,pines00,chubukov01}

The RPA approximation 
(\ref{eq: RPA chi in fermiology})
has been improved in three ways.
First, 
the single-band Hubbard model can be generalized to the three-band
Hubbard model.%
\cite{si93,zha93,liu95,millis96,kao00}
Second,
the feedback effect of the magnetic fluctuations
encoded by 
Eq.\ (\ref{eq: RPA chi in fermiology})
on the propagator of the fermionic (BCS) quasiparticles
can be included selfconsistently through the so-called
fluctuation exchange (FLEX) approximation.%
\cite{pao95,dahm96,takimoto98,dahm98,kuroki99,manske01}
Third,
all unaccounted for interactions among the fermionic (BCS) quasiparticles
can be included in a phenomenological way by substituting in
Eq.\ (\ref{eq: RPA chi in fermiology})
$\omega$ by $\omega+{i}\Gamma$ 
with $\Gamma$ a positive function of $\omega$, temperature $T$,
doping $x$, and high-$T^{\ }_c$ superconducting family.%
\cite{littlewood93,ito04}

The goal of this paper is to assess the effects 
on the incommensurate peaks of the magnetic susceptibility 
$\chi$
caused by the presence of a higher \mbox{$d$-wave} 
harmonic in the superconducting 
order parameter of underdoped Bi2212.
Indeed, it is observed in several angle resolved photoemission 
(ARPES) experiments on underdoped Bi2212 
that there are deviations from a pure \mbox{$d$-wave} order 
parameter.\cite{harris96,mesot99,borisenko02}
These deviations result in a rounding of the superconducting gap
in the vicinity of its nodes.
\cite{higher harmonics not caused by disorder}
We interpret these deviations as the signature of a
\mbox{$d$-wave} order parameter extending
to next-nearest-neighbor bonds between planar Cu sites,
i.e., they push closer to~0 
the anisotropy ratio
$v^{\ }_{\Delta}/v^{\ }_{F}$ 
between the slope of the superconducting gap
$v^{\ }_\Delta$ and the Fermi velocity $v^{\ }_F$, which are tangent and
perpendicular to the Fermi surface, respectively.

Below, we adopt a phenomenological fermiology scenario 
by which we compute the RPA susceptibility (\ref{eq: RPA chi in fermiology})
assuming that $\Pi(\omega,\mathbf{q})$ is calculated
with the BCS dispersion measured by ARPES
while
$g(\omega,\mathbf{q})=\Pi(\omega,\mathbf{q})$
and 
$h(\omega,\mathbf{q})=-U$.
To simplify the matter and to isolate the effect of
decreasing
$v^{\ }_{\Delta}(x)/v^{\ }_{F}(x)$ 
with underdoping, i.e., decreasing $x$, we treat residual lifetime effects 
on the BCS quasiparticles with the substitution
\begin{eqnarray}
\omega\longrightarrow\omega+{i}\Gamma.
\label{eq: def Gamma}
\end{eqnarray}
For reference $\Gamma=1\,meV$
is of the order of the energy resolution of most experiments.
The lowest spectral edge in
$\chi^{\prime\prime}_{\mathrm{RPA}}(\omega,\mathbf{q})$
considered as a function of $\omega$ with $\mathbf{q}$ held fixed
defines the \mbox{$\mathbf{q}$-dependent} spin gap.
We find that increasing the higher harmonics by an increment of 
15$\%$ relative to optimal doping, which corresponds to a decrease
of $T^{\ }_c$ by $28\%$ in Bi2212
as was observed in Ref.\ \onlinecite{mesot99},
decreases the \mbox{$\mathbf{q}$-dependent} spin gap 
by an amount that depends on the wave 
vector
(see Figs.~\ref{Fig: Frequency leading edge of chi''RPA moving to left with underdoping}
and~\ref{Fig: Full w-q scans})
and agrees qualitatively with YBCO measurements.
This result is a robust feature of all fermiology
scenarios and is consistent with the experimental observation 
of a fast decreasing (or even vanishing) spin gap with underdoping.
\cite{dai01,lee00}

The paper is organized as follows.
The fermiology approach is defined through a RPA magnetic susceptibility
in Sec.\
\ref{sec: RPA magnetic susceptibility}.
Numerical results are presented in Sec.\ 
\ref{sec: Numerical results}
and interpreted in Sec.~\ref{sec: Discussion}.
We close with a summary in Sec.~\ref{sec: Summary}.

\section{RPA magnetic susceptibility}
\label{sec: RPA magnetic susceptibility}

Perhaps the most important lesson that can be inferred from ARPES data
is that the low energy excitations in the superconducting state of
Bi2212 are sharply defined quasiparticles obeying the BCS
dispersion\cite{footnote on bilayer issue}
\begin{subequations}
\label{eq: def BCS dispersion from ARPES}
\begin{eqnarray}
E^{\ }_{\mathbf{k}}=
\sqrt{
\varepsilon^{2 }_{\mathbf{k}}
+
\Delta^{2 }_{\mathbf{k}}
     }
\end{eqnarray}
in the close vicinity to the Fermi surface 
defined by the condition
\begin{eqnarray}
\varepsilon^{\ }_{\mathbf{k}}=0
\label{eq: def FS}
\end{eqnarray}
and with a superconducting gap consistent with a \mbox{$d$-wave}  
symmetry, i.e.,
\begin{eqnarray}
&&
 \Delta^{\ }_{\mathbf{k}}=
-
\Delta^{\ }_{\mathbf{k}^{\prime}},
\quad
\left( k^\prime_x, k^\prime_y \right) 
=
\left( \pm  k^{\ }_y, \mp k^{\ }_x \right),
\nonumber
\\&&\\
&&
 \Delta^{\ }_{\mathbf{k}}=
\hphantom{-}
\Delta^{\ }_{\mathbf{k}^{\prime}},
\quad
\left( k^\prime_{x}, k^\prime_{y} \right)
= 
\left( \pm k^{\ }_{x}, \mp k^{\ }_{y} \right).
\nonumber
\end{eqnarray}
This fact suggests that it might be plausible  to treat the superconducting
state of high-$T^{\ }_c$ superconductors as a conventional BCS superconductor
in the close vicinity to the Fermi surface,
an assumption that we will make from now on. Following
Norman in Ref.\ \onlinecite{norman00} the measured Fermi surface is fitted from
the tight-binding expansion
\begin{eqnarray}
\varepsilon^{\ }_{\mathbf{k}}=
\frac{1}{2}
\sum_{j=0}^\infty
t^{\ }_j
\left(
\cos\mathbf{a}^{\ }_j\cdot\mathbf{k}
+
\cos\mathbf{b}^{\ }_j\cdot\mathbf{k}
\right)
\end{eqnarray}
with $\mathbf{a}^{\ }_0=\mathbf{b}^{\ }_0=0$ and
$\mathbf{a}^{\ }_j,\mathbf{b}^{\ }_j$
a pair of orthogonal vectors joining a site of the square lattice to 
two of its $j$-th nearest neighbors. Similarly, 
the \mbox{$d$-wave}  BCS gap is
fitted from the tight-binding expansion
\begin{eqnarray}
\Delta^{\ }_{\mathbf{k}}
&=&
\Delta^{\ }_1 \left( \cos k_x - \cos k_y \right)
\nonumber\\
& &+ \Delta^{\ }_2 \left( \cos 2 k_x - \cos 2 k_y \right)
\\
& &+ \Delta^{\ }_3 \left( \cos 2 k_x \cos k_y - \cos k_x \cos 2 k_y \right)
\nonumber\\
& &+ \ldots 
. \nonumber
\end{eqnarray}
\end{subequations}
In practice we truncate the expansions at $j=5$ for the Fermi surface
and to the first two terms for the \mbox{$d$-wave} BCS gap.
As opposed to the slave-boson approach which attempts to capture the
doping dependence of the $t^{\ }_j$'s and the $\Delta^{\ }_j$'s
from the large $U$-limit of the Hubbard model,
we will take the phenomenological point of view that the BCS dispersion
(\ref{eq: def BCS dispersion from ARPES})
of the ``bare'' quasiparticles
is an input deduced from ARPES data for the Bi2212 family. 

We will assume in this paper that anomalous features 
at energy scales far away from the Fermi energy, say of the order
$2\max^{\ }_{\mathbf{k}}\,\Delta^{\ }_{\mathbf{k}}$
such as is the case for the incommensurate and commensurate peaks
observed at optimal doping with inelastic neutron scattering,
can be accounted for by postulating the existence of ``bare''
noninteracting BCS quasiparticles whose dispersion is given by
a fit to ARPES data and which interact weakly through
some small residual
interactions which we take to be an on-site Hubbard repulsion with coupling
constant $U$.
In this spirit, the renormalization of the ``bare'' magnetic susceptibility 
\begin{subequations}
\label{eq: RPA magnetic susceptibility in fermiology}
\begin{eqnarray}
&&
\Pi(\omega,\mathbf{q})\!=\!
\frac{1}{N}\!
\sum_{\mathbf{k}}
\sum_{s',s=\pm}
\frac{
C^{s',s}_{\mathbf{q},\mathbf{k}}
\left[
f\bigr(s'E^{\ }_{\mathbf{k}+\mathbf{q}}\bigl)
-
f\bigr(s E^{\ }_{\mathbf{k}           }\bigl)
\right]
     }
     {
\sigma(\omega+{i}0^{+})
-
\bigl(
s'E^{\ }_{\mathbf{k}+\mathbf{q}}
-
s E^{\ }_{\mathbf{k}           }
\bigr)
     },
\nonumber\\
&&
\sigma=
\hbox{$-$ if $s'=s=-1$, $+$ otherwise},
\label{eq: ``bare'' Pi}
\\
&&
C^{s',s}_{\mathbf{q},\mathbf{k}}=
\frac{1}{4}
\left(
1
+
s's
\frac{
\varepsilon^{\ }_{\mathbf{k}+\mathbf{q}}
\varepsilon^{\ }_{\mathbf{k}           }
+
\Delta^{\ }_{\mathbf{k}+\mathbf{q}}
\Delta^{\ }_{\mathbf{k}           }
     }
     {
E^{\ }_{\mathbf{k}+\mathbf{q}}
E^{\ }_{\mathbf{k}           }
     }
\right),
\nonumber
\end{eqnarray}
is, within the RPA,
\begin{eqnarray}
\chi^{\ }_{\mathrm{RPA}}(\omega,\mathbf{q})=
\frac{
\Pi(\omega,\mathbf{q})
     }
     {
1
-
U \,
\Pi(\omega,\mathbf{q})
     }.
\label{eq: RPA chi}
\end{eqnarray}
\end{subequations}
Here,
$N$ 
is the number of sites on a square lattice with periodic boundary conditions,
$f(x)=[\exp(\beta x)+1]^{-1}$ 
is the Fermi distribution 
($\beta$ the inverse temperature),
and the BCS dispersion 
(\ref{eq: def BCS dispersion from ARPES}) 
is used. 

\begin{figure}[t] 
\includegraphics[width=0.5\textwidth]{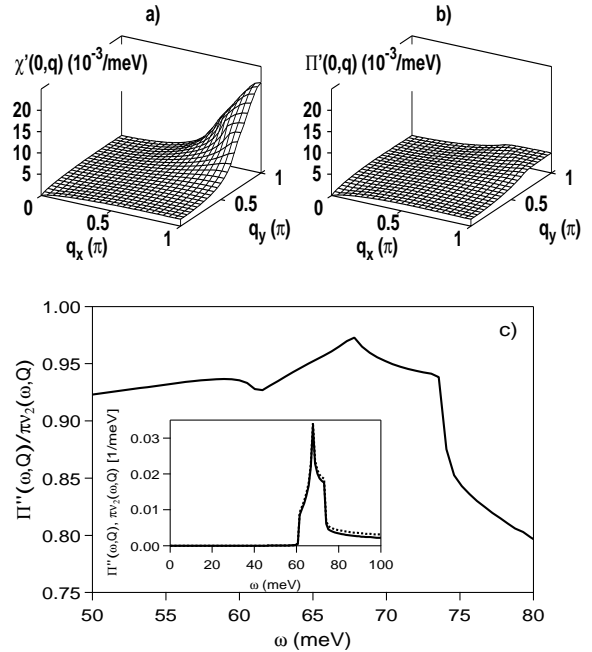}
\begin{center}
\caption{
Plot of the real parts
$\chi^{\prime}_{\mathrm{RPA}}(0,\mathbf{q})$ (a)
and 
$\Pi^{\prime}(0,\mathbf{q})$ (b)
as a function of $\mathbf{q}$ 
in the upper-right quadrant of the Brillouin zone
for the BCS dispersion of Bi2212 at optimal doping 
(see table \ref{table: BCS dispersion parameters from ARPES}).
(c)~Plot of 
$
\Pi^{\prime\prime}(\omega,\pi,\pi)
/
[\pi\nu^{\ }_2(\omega,\pi,\pi)]
$
as a function of $\omega>0$ 
in the relevant energy range 
for the BCS dispersion of Bi2212 at optimal doping 
(see table \ref{table: BCS dispersion parameters from ARPES}).
The inset displays the 
$\omega$ 
dependence of the imaginary part
$\Pi^{\prime\prime}(\omega,\pi,\pi)$
(thick line) and
$\pi\nu^{\ }_2(\omega,\pi,\pi)$
(dashed line).
The temperature is here taken to be $T=0K$ while a damping
$\Gamma=0.1 meV$ is used. 
        }
\label{Fig: static limit of chiRPA minus Pi}
\end{center}
\end{figure}

In Fig.~\ref{Fig: static limit of chiRPA minus Pi}a and 
Fig.~\ref{Fig: static limit of chiRPA minus Pi}b  we present a 
plot of the static limit of the RPA magnetic susceptibility~(\ref{eq: RPA chi})
 and 
of the static ``bare'' magnetic susceptibility
(\ref{eq: ``bare'' Pi}), respectively.
It is seen that there are no significant differences between
the static limits of the ``bare'' and RPA magnetic susceptibilities
for small momentum transfer $\mathbf{q}$ (small forward scattering)
but that for large momentum transfer, $\mathbf{q}$ close to $(\pi,\pi)$,
the renormalization effects are of order 1 as they signal the proximity
to an antiferromagnetic instability of the BCS ground state.
These renormalization effects modify the                  
quasiparticle self-energy and have been proposed as an explanation to         
the peak-dip-hump shape seen in the ARPES signal as a function 
of frequency.\cite{li00,chubukov04}
In this paper we do not consider this feedback on the quasiparticle self-energy.
However, since the effects of higher \mbox{$d$-wave} harmonics on the 
RPA magnetic susceptibility are robust, we expect them to survive in a selfconsistent
approach.

We close this section by noting that the 
imaginary part 
$\Pi^{\prime\prime}(\omega,\mathbf{q})$
of the ``bare'' susceptibility 
(\ref{eq: ``bare'' Pi})
simplifies to
\begin{eqnarray}
&&
\Pi^{\prime\prime}(\omega,\mathbf{q})=
\frac{\pi}{N}
\sum_{\mathbf{k}}
C^{+,-}_{\mathbf{q},\mathbf{k}}
\delta
\bigl(
\omega
-
E^{\ }_2(\mathbf{q},\mathbf{k})
\bigr),
\nonumber\\
&&
C^{+,-}_{\mathbf{q},\mathbf{k}}=
\frac{1}{4}
\left(
1
-
\frac{
\varepsilon^{\ }_{\mathbf{k}+\mathbf{q}}
\varepsilon^{\ }_{\mathbf{k}           }
+
\Delta^{\ }_{\mathbf{k}+\mathbf{q}}
\Delta^{\ }_{\mathbf{k}           }
     }
     {
E^{\ }_{\mathbf{k}+\mathbf{q}}
E^{\ }_{\mathbf{k}           }
     }
\right),
\label{eq: ``bare'' Pi'' at T=0}
\\
&&
E^{\ }_2(\mathbf{q},\mathbf{k})=
E^{\ }_{\mathbf{k}+\mathbf{q}}
+
E^{\ }_{\mathbf{k}           },
\nonumber
\end{eqnarray}
in the zero temperature limit and
for positive frequencies as is relevant in inelastic neutron scattering.
We shall see in Sec.\ \ref{sec: Discussion}
that the numerical results from Sec.\ \ref{sec: Numerical results}
can be understood in terms of the frequency dependence of the
2-particle density of states~(DOS)
\begin{eqnarray}
\nu^{\ }_2(\omega,\mathbf{q}):=
\frac{1}{2N}
\sum_{\mathbf{k}}
\delta
\bigl(
\omega
-
E^{\ }_2(\mathbf{q},\mathbf{k})
\bigr)
\label{eq: def nu2}
\end{eqnarray}
since it
differs very little from 
that of $\Pi^{\prime\prime}(\omega,\mathbf{q})/\pi$
for a fixed wave vector $\mathbf{q}$ 
close to the commensurate vector $(\pi,\pi)$
and at very low temperatures as is illustrated in 
Fig.\ \ref{Fig: static limit of chiRPA minus Pi}c.

\section{Numerical results}
\label{sec: Numerical results}

\begin{table}[t]
\caption{
\label{table: BCS dispersion parameters from ARPES}
Values of the hopping and gap parameters used to fit
the Fermi surface (as in Ref.~\onlinecite{norman00}) and 
gap (see Ref.~\onlinecite{mesot99}) in the Bi2212 family measured with ARPES 
as a function of doping. 
The gap parameters for $T^{\ }_c = 87K$, $T^{\ }_c = 83K$  
and $T^{\ }_c = 75K$ follow from a fit to ARPES data 
(Ref.~\onlinecite{mesot99}), 
and for $T^{\ }_c = 68K$ from an extrapolation.
The doping decreases when reading the columns from left to right:
The second, third, fourth, and fifth columns correspond to 
optimally doped, slightly underdoped, 
and underdoped samples, respectively, 
as indicated by their decreasing $T^{\ }_c$.
The critical temperatures are in units of Kelvins
while all other quantities are in units of $meV$.
        }
\begin{ruledtabular}
\begin{tabular}{crrrr}
$\hbox{Parameters}$
&
$T^{\ }_c=87K$
&
$T^{\ }_c=83K$
&
$T^{\ }_c=75K$
&
$T^{\ }_c=68K$
\\\hline\\
$t^{\ }_0$
&
$87.9$
&
$87.9$
&
$87.9$
&
$87.9$
\\
$t^{\ }_1$
&
$-554.7$
&
$-554.7$
&
$-554.7$
&
$-554.7$
\\
$t^{\ }_2$
&
$132.7$
&
$132.7$
&
$132.7$
&
$132.7$
\\
$t^{\ }_3$
&
$13.2$
&
$13.2$
&
$13.2$
&
$13.2$
\\
$t^{\ }_4$
&
$-184.9$
&
$-184.9$
&
$-184.9$
&
$-184.9$
\\
$t^{\ }_5$
&
$26.5$
&
$26.5$
&
$26.5$
&
$26.5$
\\
$\Delta^{\ }_1$
&
$18.3$
&
$18.8$
&
$18.8$
&
$19.6$
\\
$\Delta^{\ }_2$
&
$-2.1$
&
$-3.9$
&
$-5.8$
&
$-8.8$
\\
$\Delta^{\ }_0$
&
$35.0$
&
$37.0$
&
$38.0$
&
$41.0$
\\
$B$
&
$0.96$
&
$0.92$
&
$0.89$
&
$0.84$
\\
$\left.\frac{d\widetilde\Delta}{d\phi}\right|_{\phi=\pi/4}$
&
$57.0$
&
$52.0$
&
$44.8$
&
$36.3$
\\
$U$
&
$165$
&
$173$
&
$180$
&
$191$
\\
\end{tabular}
\end{ruledtabular}
\end{table}

We have calculated numerically the RPA 
(\ref{eq: RPA chi})
to the magnetic susceptibility 
with the inclusion of lifetime effects implemented through
substitution~(\ref{eq: def Gamma}).
Hereby, we are using the BCS dispersion 
(\ref{eq: def BCS dispersion from ARPES})
with the hopping and superconducting 
parameters extracted from fits to the Fermi surface and gap measured
with ARPES within the Bi2212 family 
for the underdoped and optimally doped regimes
(see Refs.~\onlinecite{norman95, norman00,mesot99,ding97,kordyuk02}).
These parameters can be found in 
table~\ref{table: BCS dispersion parameters from ARPES}.
For each doping concentration the value of the on-site Hubbard repulsion 
$U$ is chosen so that the energy of the resonance is consistent with
unpolarized neutron data in Refs.~\onlinecite{fong99,mesot00}
and with the indirect determination of the
resonance energy via the peak-dip-hump feature from APRES data
in Refs.~\onlinecite{campuzano99,zasadzinski02}.
The value of the coupling constant~$U$ 
increases with underdoping, a trend consistent with
the naive expectation that moving away 
from half-filling in the Hubbard model reduces the effect
of the strong interactions.  ARPES experiments
on Bi2212 indicate that the Fermi surface is (if at all) 
only weakly dependent on doping.\cite{ding97, kordyuk02}
 To simplify matters, we have
implemented this observation by keeping the hopping parameters unchanged as
a function of doping.

\begin{figure}[th!]
\includegraphics[angle=-0, width=0.5\textwidth]{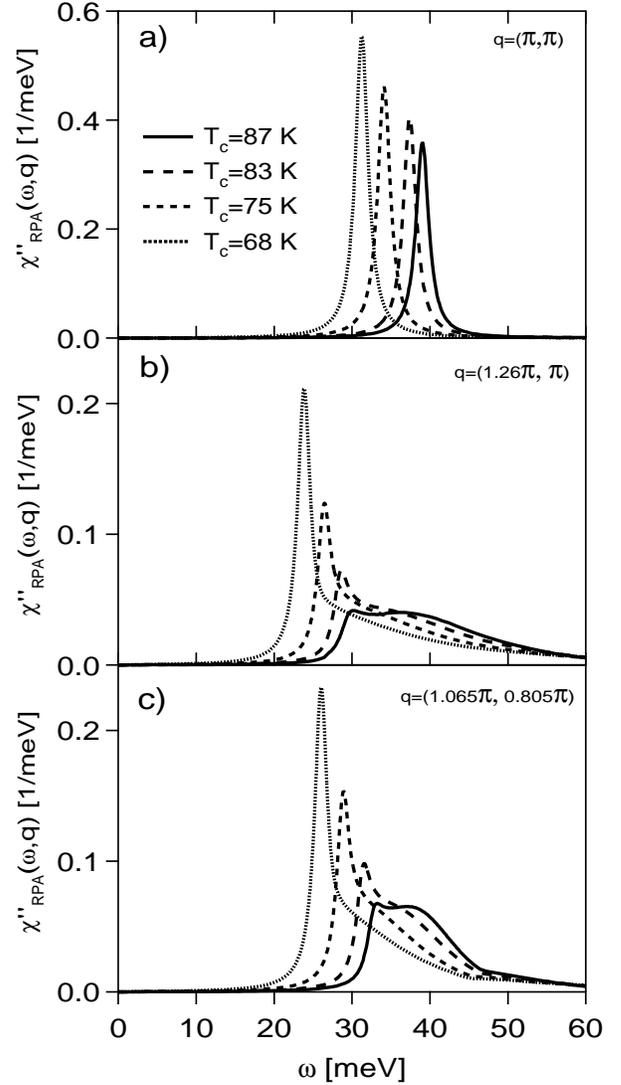}
\begin{center}
\caption{
Panel (a) displays the frequency dependences of
$\chi^{\prime\prime}_{\mathrm{RPA}}(\omega,\mathbf{q})$
at the commensurate wave vector 
$\mathbf{q}=(\pi,\pi)$
for four different values $x$ of doping in the 
Bi2212 family 
with the BCS dispersions taken from
table \ref{table: BCS dispersion parameters from ARPES}.
Curves with the leading edge moving to the left 
(i.e., to lower frequencies) have
decreasing doping $x$ ($T^{\ }_c$).
Panels (b) and (c) are the same as panel (a) except for
the incommensurate wave vectors
$\mathbf{q}=(1.26\pi,\pi)$ 
and 
$\mathbf{q}=(1.065\pi,0.805\pi)$
being held fixed.
For all three wave vectors the \mbox{$\mathbf{q}$-dependent} spin gap 
decreases with underdoping.
The temperature is here taken to be $T=5K$ while a damping
$\Gamma=1 meV$ is used. 
        }
\label{Fig: Frequency leading edge of chi''RPA moving to left with underdoping}
\end{center}
\end{figure}

To perform the summations in formula~(\ref{eq: RPA chi}), we meshed the Brillouin
zone with 
$1024\times1024$ points. 
In order to reduce the effects of a finite lifetime in 
Fig.~\ref{Fig: static limit of chiRPA minus Pi},
we reduced
$\Gamma$ in Eq.\ (\ref{eq: def Gamma}) 
from the experimental resolution $\sim 1\, meV$
to
$0.1\, meV$.
The temperature is taken to be vanishing in most instances except in
Fig.~\ref{Fig: Frequency leading edge of chi''RPA moving to left with underdoping}
and Fig.~\ref{Fig: U constant. No higher harmonics}
where $T=5K$.

\begin{figure}[t]
\includegraphics[width=0.5\textwidth, clip=false]{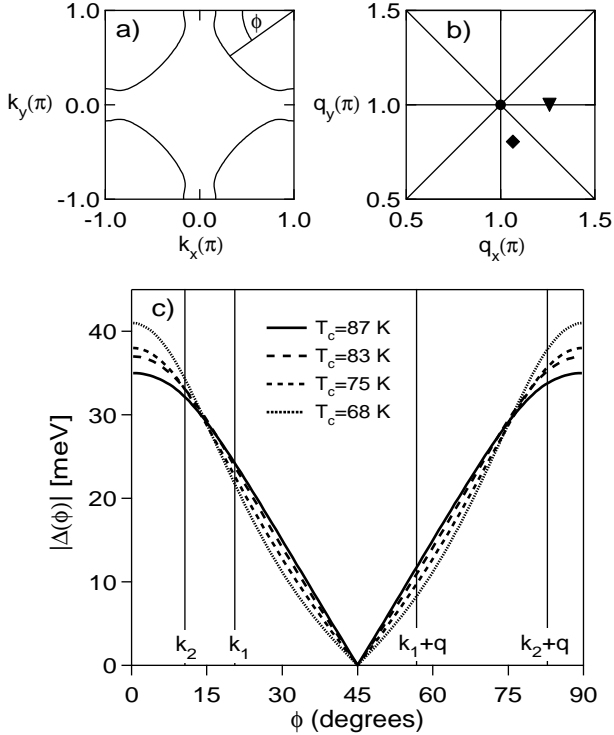}
\begin{center}
\caption{
Panel (a) displays the
Fermi surface of 
Bi2212 with  
$t_0,\ldots,t_5$ 
as in table \ref{table: BCS dispersion parameters from ARPES}.
The angle 
$\phi$
measures the position along the Fermi arc centered about 
$(\pi,\pi)$ (hole-like)
in the Brillouin zone.
Panel (b)
displays the reciprocal
\mbox{$\mathbf{q}$ space} together with 
the symmetry axes (parallel and diagonal) and the three 
wave vectors 
$(\pi,\pi)$,
$(1.26\pi,\pi)$, 
and $(1.065\pi,0.805\pi)$.
Panel (c) displays the absolute value of the gap function
$|{\Delta}(\phi)|$
on the Fermi surface as a function of Fermi surface angle~$\phi$
for 
Bi2212 
with  
the parameters of table
\ref{table: BCS dispersion parameters from ARPES}.
Plotted are the gaps for optimally doped samples (solid line), 
slightly underdoped (long dashed line), 
and underdoped samples (dashed and dotted lines).
For the momentum transfer
$\mathbf{q} = (1.26\pi,\pi)$
vertical lines at the Fermi surface angles,
where the pairwise nested wave vectors
$\mathbf{k}_1, \mathbf{k}_1 + \mathbf{q}, \mathbf{k}_2$ 
and
$\mathbf{k}_2 + \mathbf{q}$ 
lie, are drawn (see Sec.\ \ref{sec: Discussion}).
        }
\label{Fig: FS of Bi2212}
\end{center}
\end{figure}

It is convenient to use polar coordinates to represent the 
wave-vector dependence 
of the superconducting gap on the Fermi surface. To this end, define
along the arc of the Fermi surface (\ref{eq: def FS})
that belongs to the upper-right quadrant of the Brillouin zone 
the angle $0\leq\phi\leq\pi/2$ through 
(see Fig.~\ref{Fig: FS of Bi2212}a)
\begin{eqnarray}  
\phi = 
\mathop{\mathrm{arctan}}\left( \frac{\pi - k_y}{\pi - k_x}\right).
\label{eq: FS angle}
\end{eqnarray}
In terms of this angle, define the auxiliary gap function
\begin{eqnarray}
&&
\widetilde{\Delta}(\phi):=
\Delta^{\ }_0
\left[
B\cos(2\phi)
+
\left(1-B\right) \cos(6\phi) 
\right],
\nonumber\\
&&
\Delta^{\ }_0:=
\max_{\mathbf{k}\in\hbox{\tiny Fermi surface}}
\Delta^{\ }_{\mathbf{k}}.
\label{eq: auxilliary gap of phi}
\end{eqnarray}
The auxiliary gap function 
$\widetilde{\Delta}(\phi)$ 
obeys the same \mbox{$d$-wave} symmetry as the gap function
$\Delta^{\ }_{\mathbf{k}}$
does on the Fermi surface.
The parameter $B$ entering  
$\widetilde{\Delta}(\phi)$
is taken to be a function of doping 
$x$ that can be determined for Bi2212
from ARPES measurements 
(see table \ref{table: BCS dispersion parameters from ARPES}).%
\cite{mesot99}
The agreement between $\widetilde\Delta(\phi)$ and $\Delta^{\ }_{\mathbf{k}}$
for $\mathbf{k}$ on the Fermi surface is excellent.
The Fermi velocity $v^{\ }_{\Delta}$
tangent to the Fermi surface at the node
of the gap is proportional to 
$(d\widetilde\Delta)/(d\phi)$
at
${\phi=\pi/4}$.
It decreases with underdoping as is apparent from table
\ref{table: BCS dispersion parameters from ARPES}.
The 
gap function
$\Delta_\mathbf{k}$ 
on the Fermi surface is depicted in
Fig.~\ref{Fig: FS of Bi2212}c 
for the parameters of table \ref{table: BCS dispersion parameters from ARPES}. 

In Figs.\ 
\ref{Fig: Frequency leading edge of chi''RPA moving to left with underdoping}a-c
we illustrate
for three different wave vectors depicted in Fig.\ \ref{Fig: FS of Bi2212}b,
 $\mathbf{q}=(\pi,\pi)$, the antiferromagnetic wave vector,
$\mathbf{q}=(1.26\pi,\pi)$, a wave vector on the horizontal 
symmetry axis passing through $(\pi,\pi)$,
and 
$\mathbf{q}=(1.065\pi,0.805\pi)$, a wave vector off the symmetry axes,
 how the dependences on frequency 
of the imaginary part
$\chi^{\prime\prime}_{\mathrm{RPA}}(\omega+i\Gamma,\mathbf{q})$
of the RPA magnetic susceptibility (\ref{eq: RPA chi}) changes
with underdoping.
In each panel the four curves correspond to the four dopings
in table~\ref{table: BCS dispersion parameters from ARPES}.
As the doping decreases (i.e., decreasing $T^{\ }_c$)
the leading edge moves to the left (i.e., to lower frequencies).
The \mbox{$\mathbf{q}$-dependent} spin gap, 
which we define by the lowest spectral edge in 
 $\chi^{\prime \prime}_{\mathrm{RPA}}$
considered as a function of frequency, 
decreases with decreasing doping
for both the commensurate wave vector $(\pi,\pi)$ and the 
incommensurate points
$(1.26\pi,\pi)$ and $(1.065\pi,0.805\pi)$
on the symmetry axis and off the symmetry axis passing through~$(\pi,\pi)$,
respectively. This behavior is consistent with the observed doping dependence
of the \mbox{$\mathbf{q}$-dependent} spin gap in the LSCO and YBCO families.
The intensities of the peaks in all panels of Fig.\
\ref{Fig: Frequency leading edge of chi''RPA moving to left with underdoping}a-c
increase with underdoping. This increase in intensity is accompanied by
a narrowing of the line shape in panels b and c.

\begin{figure}[t]
\includegraphics[width=0.45\textwidth]{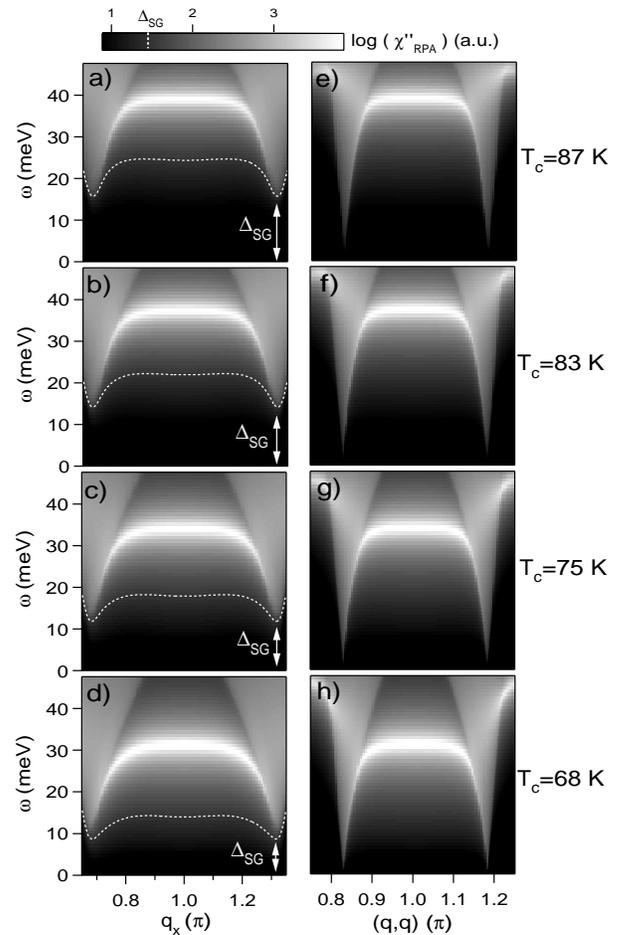} 
\begin{center}
\caption{
Panels (a)-(d) display the intensities 
of~$\chi^{\prime\prime}_{\mathrm{RPA}}(\omega,q_x,\pi)$
in the $(\omega,q_x)$ plane at fixed $q_y=\pi$ for four different doping 
concentrations 
in the Bi2212 family with the BCS dispersions taken from
table~\ref{table: BCS dispersion parameters from ARPES}.
Panels (a), (b), (c), and (d) correspond to samples with a transition 
temperature $T_c$ of
$87K$, $83K$, $75K$, and $68K$, respectively. As a guide to the eye 
contour lines are drawn at an intensity which is about 0.5\% 
of the maximum intensity of the dynamical spin susceptibility.
Panels (e)-(h) are the same as panels (a)-(d) except for
the~\mbox{$\bf{q}$-vector} which runs along
the diagonal instead of the horizontal line in reciprocal space.
The temperature is here taken to be $T=0K$ while a damping
$\Gamma=1 meV$ is used.}
\label{Fig: Full w-q scans}
\end{center}
\end{figure}

In Fig.~\ref{Fig: Full w-q scans} the intensities of the imaginary
part of the RPA spin susceptibility are shown as a function of frequency and 
wave vector for the four doping concentrations of 
table~\ref{table: BCS dispersion parameters from ARPES}. 
At the corresponding resonance energies 
($39meV$, $37meV$, $34meV$, and $31meV$, respectively) 
the peaks are at the $(\pi, \pi)$ point. 
When the frequency is reduced from the resonance energy, 
dominant incommensurate peaks in the 
$(\pi + \delta, \pi)$ and $(\pi, \pi + \delta)$ direction occur,
and a subdominant structure on the diagonal lines 
which pass through two incommensurate points in $\mathbf{q}$ space shows up.
The peaks are dispersing with a downward curvature. 
Above the resonance energy 
the dominant peaks cross over from the parallel to the diagonal
symmetry axes passing through $(\pi,\pi)$.
The same happens below
$19.2meV$ ($T_c=87K$), $17.6meV$ ($T_c = 83K$), $15.2meV$ ($T_c=75K$),
and $12.8meV$ ($T_c=68K$), respectively.
(We will then speak of parallel and diagonal peaks, respectively.)
For frequencies larger than the resonance energy incommensurate peaks 
on the diagonal symmetry axes passing through
the antiferromagnetic wave vector $(\pi, \pi)$ 
have recently been observed in both La$_{2-x}$Ba$_x$CuO$_4$\cite{tranquada04}
and YBCO.\cite{hayden04}
The crossover to dominant diagonal peaks at low energies has not been 
observed in experiments.
If the wave vector is restricted to the horizontal 
symmetry axes in the Brillouin zone,
the spin gap~$\Delta^{\ }_{\mathrm{SG}}(q^{\ }_x,\pi)$ 
remains finite for all values of
$q^{\ }_x$. 
However, if the wave vector lies on the diagonal
symmetry axes the spin gap ~$\Delta^{\ }_{\mathrm{SG}}(q^{\ }_x,q^{\ }_x)$ 
vanishes for certain values
of $q^{\ }_x=q^{\ }_y$.

\begin{figure}[t]
\includegraphics[angle=-0, width=0.5\textwidth]{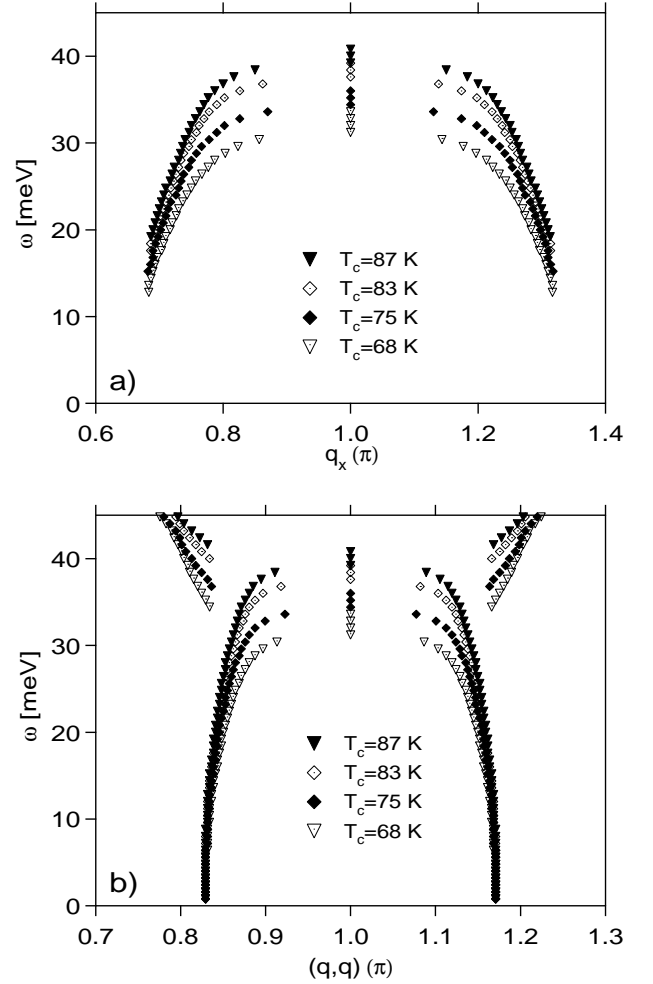}
\begin{center}
\caption{
(a) Peak positions of 
$\chi^{\prime\prime}_{\mathrm{RPA}}(\omega,\mathbf{q})$
for fixed $q_y=\pi$ as a function of
$q_x$ and  energy
with
the BCS dispersions taken from
table~\ref{table: BCS dispersion parameters from ARPES} for
four different samples with a transition temperature $T_c$ of
$87K$ (solid triangles), $83K$ (open diamonds), $75K$ (solid diamonds), 
and $68K$ (open triangles). 
(b) Peak positions of 
$\chi^{\prime\prime}_{\mathrm{RPA}}(\omega,\mathbf{q})$
as function of frequency and momentum transfer with $\mathbf{q}$ along 
the diagonal $q_x=q_y$ 
for the same four doping concentrations as in panel (a). 
Above the resonance 
the diagonal peaks are the dominant structure. 
The same happens at very low energies.
At intermediate energies, 
the diagonal peaks
 are subdominant with the parallel peaks being dominant 
(see Fig.~\ref{Fig: Full w-q scans}).
The temperature is here taken to be $T=0K$ while a damping
$\Gamma=1 meV$ is used. 
        }  
\label{Fig: Peak position of imaginary part of RPA susceptibility}
\end{center}
\end{figure}

The dispersion of the peak positions of 
$\chi^{\prime \prime}_{\mathrm{RPA}}$ 
for the four doping concentrations of 
table~\ref{table: BCS dispersion parameters from ARPES}
and with 
$\mathbf{q}$ moving along the horizontal or diagonal 
lines passing through 
$(\pi, \pi)$  are shown in 
Fig.~\ref{Fig: Peak position of imaginary part of RPA susceptibility}a and
Fig.~\ref{Fig: Peak position of imaginary part of RPA susceptibility}b, 
respectively.
With underdoping the bell-like shape of the peak dispersion moves 
to lower energies implying a decrease of the 
\mbox{$\mathbf{q}$-dependent} spin gap 
(see also Figs.~%
\ref{Fig: Frequency leading edge of chi''RPA moving to left with underdoping}
and~\ref{Fig: Full w-q scans}).
The degree of incommensuration
measured by the separation $2\delta$ of the two peaks of
$\chi^{\prime\prime}_{\mathrm{RPA}}$ 
decreases as one moves from the optimally
doped to the underdoped regimes.
This agrees with neutron scattering measurements on the YBCO and LSCO 
families.\cite{dai01,yamada98} 

\section{Discussion}
\label{sec: Discussion}

In this section we review the mechanism causing commensurate
resonance peaks and incommensurate peaks in the numerical simulations
of the imaginary part of the RPA magnetic susceptibility.
The discussion is kept brief as it has been
thoroughly described in the fermiology papers of the early 90's with
the exception of one aspect based on some point group symmetries of the BCS
dispersion.

Up to multiplication with the  Bose-Einstein distribution,
inelastic neutron scattering has direct access to the imaginary part
$\chi^{\prime\prime}(\omega,\mathbf{q})$
of the magnetic susceptibility
$\chi(\omega,\mathbf{q})$,
which is given by
\begin{eqnarray}
&&
\chi^{\prime\prime}_{\mathrm{RPA}}(\omega,\mathbf{q})=
\frac{
\Pi^{\prime\prime}(\omega,\mathbf{q})
     }
     {
\bigl[1-U\,\Pi^{\prime      }(\omega,\mathbf{q})\bigr]^2
+
\bigl[  U\,\Pi^{\prime\prime}(\omega,\mathbf{q})\bigr]^2
     }
\nonumber\\
&& \label{eq: im part of chi_RPA}
\end{eqnarray}
within the RPA approximation 
[$\chi^{\prime      }(\omega,\mathbf{q})$ 
and 
 $\Pi^{\prime      }(\omega,\mathbf{q})$ 
denote the real parts of 
$\chi(\omega,\mathbf{q})$ 
and 
$\Pi(\omega,\mathbf{q})$,
respectively]. 
The ``bare'' imaginary part 
$\Pi^{\prime\prime}(\omega,\mathbf{q})$
of the magnetic susceptibility
controls
$\chi^{\prime\prime}_{\mathrm{RPA}}(\omega,\mathbf{q})$
in two ways. First, 
$\chi^{\prime\prime}_{\mathrm{RPA}}(\omega,\mathbf{q})$
vanishes whenever 
$\Pi^{\prime\prime}(\omega,\mathbf{q})$
does and
$1 \ne U \, \Pi^{\prime      }(\omega,\mathbf{q})$.
 Second, at $\mathbf{q}$ held fixed,
any steplike discontinuity in the frequency dependence of
$\Pi^{\prime\prime}(\omega,\mathbf{q})$ 
at some frequency 
$\omega^{\ }_{{n}}(\mathbf{q})$
results in a logarithmic divergence of the frequency dependence of
$\Pi^{\prime      }(\omega,\mathbf{q})$
at
$\omega^{\ }_{{n}}(\mathbf{q})$
through the Kramers-Kronig relation between the imaginary and
real parts of causal response functions.
This in turn guarantees
that (i)~the dynamical Stoner condition
\begin{subequations}
\begin{eqnarray}
1-U \, \Pi^{\prime      }(\omega,\mathbf{q})=0
\label{eq: Stoner condition}
\end{eqnarray}
can be met at the frequency
\begin{eqnarray}
\omega^{* }_{{n}}(\mathbf{q})<\omega^{\ }_{{n}}(\mathbf{q})
\end{eqnarray}
and
(ii)~$\chi^{\prime\prime}_{\mathrm{RPA}}(\omega,\mathbf{q})$
acquires a pole at~$\omega^{* }_{{1}}(\mathbf{q})$ and peaks at 
$\omega^{* }_{{n}}(\mathbf{q})$, where $n>1$
indexes the remaining jump discontinuities of 
$\Pi^{\prime\prime}(\omega,\mathbf{q})$.

For any small but finite damping $\Gamma$,
$\Pi^{\prime\prime}(\omega+ i \Gamma,\mathbf{q})$ becomes
a nonvanishing and continuous function of 
$\omega>0$ when holding $\mathbf{q}$ fixed.
Similarly,
the logarithmic singularities in the $\omega$ dependences of
$\Pi^{\prime      }(\omega,\mathbf{q})$ 
are cutoff by a finite $\Gamma$
under the substitution~(\ref{eq: def Gamma}).
Under this substitution
the dynamical Stoner criterion 
(\ref{eq: Stoner condition})
can only be met for a sufficiently large
size of the step in $\Pi^{\prime \prime}(\omega, \mathbf{q})$,
and the pole at $\omega^{* }_{1}$  in
$\chi^{\prime\prime}_{\mathrm{RPA}}(\omega,\mathbf{q})$
turns into a peak
of finite height
\begin{eqnarray}
\chi^{\prime\prime}_{\mathrm{RPA}}(\omega^{* }_{1} + i \Gamma,\mathbf{q})=
\frac{
1
     }
     {
U^2 \,
\Pi^{\prime\prime}(\omega^{* }_{1} + i \Gamma,\mathbf{q})
     }
\end{eqnarray}
with the full width at half maximum (FWHM)
\begin{eqnarray}
\mathrm{FWHM}= 
\left.
\frac{
2
\Pi^{\prime\prime}(\omega + i \Gamma,\mathbf{q})
     }
     {
\bigl(
\partial
\Pi^{\prime      }(\omega + i \Gamma,\mathbf{q})/
\partial\omega
\bigr)
     }
\right|^{\ }_{\omega=\omega^{* }_{1}}
.
\end{eqnarray}
\end{subequations}

To sum up, the condition for dynamical Stoner enhancement of
$\chi^{\prime\prime}_{\mathrm{RPA}}(\omega,\mathbf{q})$,
a collective effect,
can be reduced to the condition for 
$\Pi^{\prime\prime}(\omega,\mathbf{q})$
to have steplike discontinuities as a function of frequency when holding
the wave vector fixed.

In view of 
Eqs.~(\ref{eq: ``bare'' Pi'' at T=0})
 and~(\ref{eq: def nu2})
steplike discontinuities in the frequency dependence of
$\Pi^{\prime\prime}(\omega,\mathbf{q})$
with $\mathbf{q}$ held fixed
(at zero temperature and with the inverse lifetime
$\Gamma$ 
infinitesimally small)
are closely related 
to steplike discontinuities in the frequency dependence of the 2-particle DOS 
(\ref{eq: def nu2})
at fixed $\mathbf{q}$.
These jump discontinuities in
$\nu^{\ }_2(\omega,\mathbf{q})$
occur whenever the 2-particle energy 
$E^{\ }_2(\mathbf{q},\mathbf{k})$ 
in Eq.\ 
(\ref{eq: ``bare'' Pi'' at T=0})
reaches a local minimum
$\omega^{\ }_{n}$
at the $\mathbf{q}$-dependent wave vector
$\mathbf{k}^{\  }_{n}$, 
in which case the size of the step is proportional 
to\cite{footnote on definition of k_1 and k_2} 
\begin{eqnarray}
&&
\left(m^{(1)}_{n} m^{(2)}_{n}\right)^{1/2},
\nonumber\\
&&
\label{eq: stepsize}
\\
&&
0<\frac{1}{m^{(1,2)}_{n}}=
\left.
\frac{
\partial^2 E^{\ }_2(\mathbf{q},\mathbf{k})
     }
     {
\partial k^{\ }_{1,2}\partial k^{\ }_{1,2}
     }
\right|^{\ }_{\mathbf{k}=\mathbf{k}^{\ }_{n}}.
\nonumber
\end{eqnarray}
The steplike discontinuity in the 2-particle DOS 
$\nu^{\ }_2(\omega,\mathbf{q})$
at 
$\omega^{\ }_{n}$
is present in
$\Pi^{\prime\prime}(\omega,\mathbf{q})$
if and only if the coherence factor
$C^{+,-}_{\mathbf{q},\mathbf{k}}$
from Eq.\ (\ref{eq: ``bare'' Pi'' at T=0})
is non-vanishing in a neighborhood
of $\mathbf{k}^{\ }_{n}$.
The steplike jump is turned into a smooth increase otherwise.%
\cite{kee99}

It is shown in appendix 
\ref{appsec: Degeneracy}
that, 
when $\mathbf{q}$ is chosen not to be on any of
the diagonals of the Brillouin zone or the horizontal and vertical
lines passing through the commensurate point
$(\pi,\pi)$ (see Fig.~\ref{Fig: FS of Bi2212}b),
there are four distinct but 2-fold degenerate local minima
$\omega^{\ }_1<\omega^{\ }_2<\omega^{\ }_3<\omega^{\ }_4$
of
$E^{\ }_2(\mathbf{q},\mathbf{k})$ 
for the parameters of table 
\ref{table: BCS dispersion parameters from ARPES}
located at the $\mathbf{q}$-dependent wave vector
$\mathbf{k}_1$,
$\mathbf{k}_2$, 
$\mathbf{k}_3$, 
and~$\mathbf{k}_4$,
respectively,
in the Brillouin zone.
When $\mathbf{q}$ is chosen
to be on the diagonals that pass through 
the commensurate point~$(\pi, \pi)$, but away from it,
there are are three distinct local 
minima~$\omega^{\ }_1<\omega^{\ }_2<\omega^{\ }_3$ at
$\mathbf{k}_1$, 
$\mathbf{k}_2$, 
and~$\mathbf{k}_3$, 
respectively, 
which are either
doubly degenerate ($\omega^{\ }_1$ and~$\omega^{\ }_3$) or
4-fold degenerate ($\omega^{\ }_2$).
If the momentum transfer lies on the horizontal
or vertical lines passing
through the commensurate point,
but away from it,
there are two distinct but 4-fold degenerate local 
minima~$\omega^{\ }_1<\omega^{\ }_2$
of~$E^{\ }_2(\mathbf{q},\mathbf{k})$
located at the $\mathbf{q}$-dependent wave vector
$\mathbf{k}_1$ 
and
$\mathbf{k}_2$, respectively.
Finally, all local minima collapse to one global but 8-fold degenerate minimum
$\omega^{\ }_1$
located at 
$\mathbf{k}^{\ }_1$
when $\mathbf{q}=(\pi,\pi)$ or one of its symmetry related images.

Two criteria control the size of the first step with increasing~$\omega$: 
(i)  How large the effective masses $m^{(1,2)}_1$ are, i.e., how
flat the 2-particle dispersion is,
(ii) how large the  degeneracy of the global minimum is.
It turns out that, when 
$\mathbf{q}$
is chosen to be at the commensurate antiferromagnetic vector 
$(\pi,\pi)$,
the geometrical mean of the effective masses $m^{(1,2)}_1$ becomes very large
because of the proximity of $\mathbf{k}^{\ }_1$ to an extended saddle-point
for the parameters of table 
\ref{table: BCS dispersion parameters from ARPES}.
Furthermore this effect is magnified by the 8-fold degeneracy of the
global minimum of the 2-particle dispersion at $\mathbf{q}=(\pi,\pi)$.
Taken together conditions (i) and (ii) insure that the 
dynamical Stoner criterion is met at an energy
\begin{eqnarray}
\omega^{* }_1\equiv \omega^{* }_{\mathrm{AF}}
\end{eqnarray}
sufficiently far below the 2-particle continuum threshold
for the resonant nature of the resulting spin-1 collective excitation
not to be washed out by
a finite temperature of $5\,K$ or a damping of $1\,meV$.
The 8-fold degeneracy is partially broken 
to a 4-fold degeneracy
when $\mathbf{q}$ moves away
from $(\pi,\pi)$ along the horizontal or vertical lines
passing through $(\pi,\pi)$ (parallel incommensuration)
with $\mathbf{k}^{\ }_1$
moving towards the nodal line 
 and $\mathbf{k}^{\ }_2$
moving away from it
(see Fig.\ \ref{Fig: FS of Bi2212}c). 
For energies below the commensurate resonance energy
\begin{eqnarray} 
\omega^{* }_1\equiv
\omega^{* }_{\parallel\mathrm{IC}}<
\omega^{* }_{\mathrm{AF}} ,
\end{eqnarray}
this turns the resonant commensurate peak into
a downward dispersing incommensurate peak, which 
eventually becomes  nonresonant
due to the failure to meet the Stoner criterion.
By energy conservation the dominant incommensurate peaks cross over to
the diagonals of the Brillouin zone at frequencies much lower than
$\omega^{* }_{\mathrm{AF}}$ 
since $\mathbf{q}$ then connects regions close to
the nodes of the BCS dispersion.\cite{lu92}
When the wave vector is pushed away from~$(\pi, \pi)$  along the diagonals, 
the \mbox{8-fold} global minimum splits into three distinct minima. 
The first minimum $\omega_1$ being lowest in energy 
leads to the diagonal peaks at 
low energies. Together 
with the lowest minimum of the 2-particle energy for 
$\mathbf{q}$ off the symmetry axes it is also the cause of 
the subdominant structure on the diagonal lines passing 
through $(\pi + \delta, \pi)$ and $(\pi, \pi + \delta)$, say,  at intermediate energies.
 The second \mbox{4-fold} degenerate  minimum lies in general at energies 
just above the resonance $\omega^*_{\mathrm{AF}}$, but below the 
\mbox{8-fold} degenerate minima at $\mathbf{q}=(\pi, \pi)$,  
and is thus responsible for the crossover to diagonal incommensurate peaks 
above the resonance energy.
Finally, by choosing $\mathbf{q}$
away from the symmetry axes one spreads the spectral weight of the
commensurate resonance among four distinct \mbox{2-fold} 
degenerate local minima, 
three of which are in the 2-particle continuum
thereby loosing the most in height and sharpness of the peak relative
to the line shape of the commensurate resonance as a function of frequency.

To conclude, 
the crossover from diagonal to parallel incommensuration 
with increasing frequency at low energies and from parallel 
to diagonal peaks with increasing frequency above the resonance energy 
is brought about by the onset of 
\mbox{4}-fold degenerate minima of the 2-particle dispersion 
$E_2(\mathbf{q}, \mathbf{k})$ 
at the corresponding energies.
This effect is quite sensitive to both the shape of the BCS dispersions
 and the details of the 
residual interaction among the BCS quasiparticles:
On the one hand, an antiferromagnetic interaction with 
$h(\omega, \mathbf{q}) = (J/2) (\cos q^{\ }_x + \cos q^{\ }_y)$
leads to an enhancement of the diagonal peaks compared to the parallel ones, 
which potentially destroys the crossover to parallel incommensuration
upon increasing frequency towards the resonance.%
\cite{norman00} 
On the other hand,  
it was observed in 
Refs.~\onlinecite{brinckmann01,li02} 
that the parallel peaks remain dominant at energies below 
but not too far from the resonance
in a slave-boson mean-field
approach to the \mbox{$t-t^\prime-J$ model}.

At low energies the intensity of 
$\chi^{\prime\prime}_{\mathrm{RPA}}(\omega+i\Gamma,\mathbf{q})$
for a fixed wave vector 
dies off quickly (see 
Figs.~%
\ref{Fig: Frequency leading edge of chi''RPA moving to left with underdoping}
and~\ref{Fig: Full w-q scans}). 
In the following we shall define
the lowest, sharp spectral edge in
$\chi^{\prime\prime}_{\mathrm{RPA}}(\omega,\mathbf{q})$
considered as a function of frequency at a fixed wave vector $\mathbf{q}$
to be the $\mathbf{q}$-dependent spin gap.
By 
Eqs.\ (\ref{eq: ``bare'' Pi'' at T=0})
and   (\ref{eq: im part of chi_RPA}), 
the $\mathbf{q}$-dependent spin gap for vanishing damping $\Gamma$ is given by 
the threshold to the 2-particle continuum provided there is no resonant 
excitation below the threshold.
Otherwise, the $\mathbf{q}$-dependent spin gap is 
determined by the energy of the lowest bound state.
Any sharp spectral edge in
$\chi^{\prime\prime}_{\mathrm{RPA}}(\omega,\mathbf{q})$
turns into a smooth edge by the inclusion of
finite lifetime effects through
substitution~(\ref{eq: def Gamma})
in which case, to a first approximation,
  the $\mathbf{q}$-dependent spin gap
 is given by the position of the half maximum of the leading
edge on the low frequency side.

The effects on the $\mathbf{q}$-dependent spin gap 
caused by the change of the gap parameters in
table~\ref{table: BCS dispersion parameters from ARPES}
induced by underdoping
can be best understood by use of arguments 
based on pseudo Fermi nesting.
A good approximation for the position 
of the \mbox{$\mathbf{q}$-dependent} (local) minima 
of~$E^{\ }_2(\mathbf{q}, \mathbf{k})$  is obtained 
by requesting that 
both~$\mathbf{k}^{\ }_{n}$ 
and~$\mathbf{k}^{\ }_{{n}}+\mathbf{q}$ 
lie on the Fermi surface
\begin{subequations}
\begin{eqnarray} \label{eq: pseudo nesting equation}
0=
\varepsilon^{\ }_{\mathbf{k}^{\ }_{n}}=
\varepsilon^{\ }_{\mathbf{k}^{\ }_{n}+\mathbf{q}}
\end{eqnarray}
and that the coherence factor $C^{+,-}_{\mathbf{q}, \mathbf{k}}$ be maximal, 
i.e.,
\begin{eqnarray}
\mathrm{sgn}
\bigl(
\Delta^{\ }_{\mathbf{k}^{\ }_{n}}
\Delta^{\ }_{\mathbf{k}^{\ }_{n}+\mathbf{q}}\bigr)=-1.
\end{eqnarray}
For a fixed wave vector~$\mathbf{q}$
the steplike discontinuity of the frequency dependence 
of $\Pi^{\prime \prime}(\omega, \mathbf{q})$
is then approximately located at the energy
\begin{eqnarray}
\widetilde{\omega}^{\ }_n 
= 
\left| \Delta(\widetilde{\mathbf{k}}^{\ }_n) \right|
+ \left| \Delta(\widetilde{\mathbf{k}}^{\ }_n + \mathbf{q} ) \right|,
\end{eqnarray}
where $\widetilde{\mathbf{k}}^{\ }_n$ are the solutions of the equation 
set~(\ref{eq: pseudo nesting equation}).
\end{subequations}
In particular, $\widetilde{\omega}^{\ }_1$ determines approximately the
threshold to the \mbox{2-particle} continuum.

At frequencies much smaller than the resonance 
energy~$\omega^{*}_{\mathrm{AF}}$ 
the approximate \mbox{2-particle} threshold~$\widetilde{\omega}_{1}$
is effectively governed by 
excitations whose pseudo-Fermi-nesting vectors 
connect parts of the Fermi surface which are
close to the nodes. Hence, at these energies the decreasing slope of the gap 
function~$(d \Delta)/(d \phi)$ at $\phi=\pi/4$
results in a decrease of the approximate \mbox{2-particle} 
threshold~$\widetilde{\omega}_1$
(see vertical lines~$\mathbf{k}_1$ and~$\mathbf{k}_1 + \mathbf{q}$ 
in Fig.~\ref{Fig: FS of Bi2212}c).

For frequencies which are comparable to the resonance energy however, 
the \mbox{2-particle} threshold increases with the change of the gap 
parameters induced by underdoping.
This can be understood from the fact that at frequencies close to, 
or higher than
the resonance energy the 
approximate \mbox{2-particle} threshold~$\widetilde{\omega}_1$ 
is controlled by excitations with
pseudo-Fermi-nesting vectors that connect parts of the Fermi surface 
which are in the vicinity 
of~$(0,\pi)$ 
or one of its symmetry related images.
In these parts of the Fermi surface
the increase of the gap maximum~$\Delta_0$ causes the gap function 
to increase with underdoping
(see vertical lines~$\mathbf{k}_2$ and~$\mathbf{k}_2 + \mathbf{q}$ 
in Fig.~\ref{Fig: FS of Bi2212}c).

\begin{figure}[th!]
\includegraphics[angle=-0, width=0.5\textwidth]{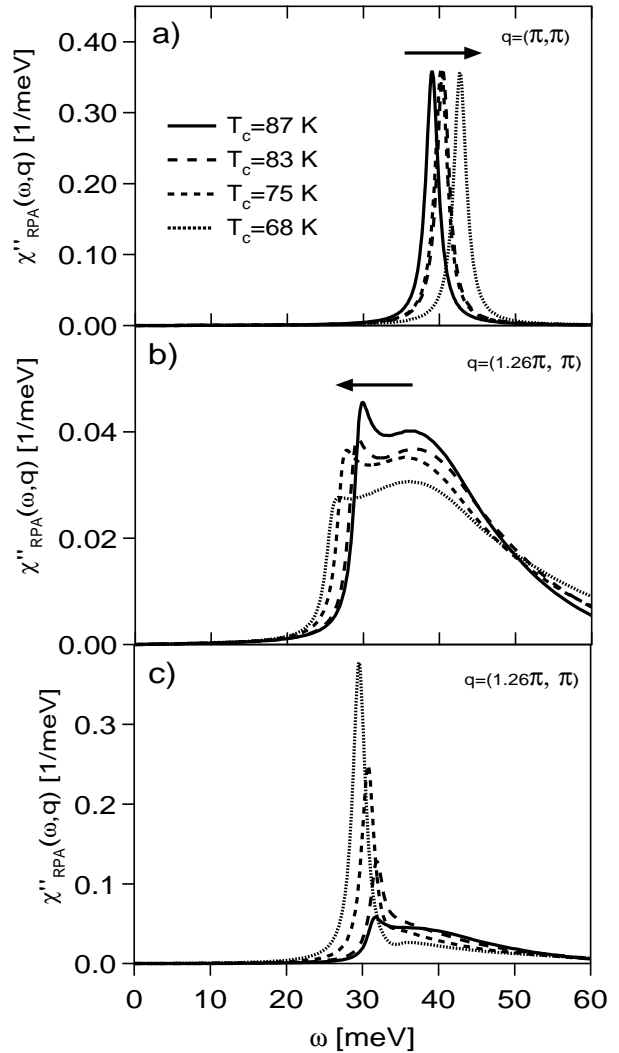}
\begin{center}
\caption{
Panel (a) displays the frequency dependences of
$\chi^{\prime\prime}_{\mathrm{RPA}}(\omega,\mathbf{q})$
at the commensurate wave vector 
$\mathbf{q}=(\pi,\pi)$
for four different values $x$ of doping in the 
Bi2212 family 
with the BCS dispersions taken from
table \ref{table: BCS dispersion parameters from ARPES}
 but with $U$ fixed to $165meV$ for all doping concentrations. 
Panel (b) is the same as panel (a) except for
the incommensurate wave vector
$\mathbf{q}=(1.26\pi,\pi)$ 
being held fixed.
Panel (c) 
displays the frequency dependences of
$\chi^{\prime\prime}_{\mathrm{RPA}}(\omega,\mathbf{q})$
at the incommensurate wave vector 
$\mathbf{q}=(1.26\pi,\pi)$ 
for four different values $x$ of doping in the 
Bi2212 family 
without inclusion of higher harmonics, i.e., with
vanishing  $\Delta^{\ }_2$.
As explained in the text we take here
$\Delta^{\ }_1$ to be $18.8meV$, $19.9 meV$, $20.5 meV$, and $22.1meV$, 
for samples with a $T_c$ of $87K$, $83K$, $75K$, and $68K$, respectively,
and $U$ increases with underdoping from
 $170meV$ at optimal doping to
$184meV$,
$196meV$,
and
$211meV$ in the underdoped regime.
In this way, $U$ is adjusted to reproduce the positions
of the resonance as observed in Bi2212.
The temperature is here taken to be $T=5K$ while a damping
$\Gamma=1 meV$ is used. 
        }
\label{Fig: U constant. No higher harmonics}
\end{center}
\end{figure}

On top of the effects caused by the change of the gap parameters 
there is an additional modification
of~$\chi^{\prime \prime}_{\mathrm{RPA}}(\omega+i\Gamma,\mathbf{q})$ 
with underdoping induced by the increase of the coupling constant~$U$. 
Owing to the failure to meet the Stoner criterion, the frequency  
of the incommensurate peaks at a given wave vector~$\mathbf{q}$ and 
at energies much smaller than the resonance energy is not changed by a 
not-too-large increase in $U$. Thus, in the limit of 
a~$\Gamma=0$, the $\mathbf{q}$-dependent spin gap 
is unaffected in this energy range by a not-too-large increase 
of the coupling constant.
For energies close to the resonance energy~$\omega^{*}_{\mathrm{AF}}$ though,
the increase of~$U$ with underdoping causes both 
the \mbox{$\mathbf{q}$-dependent} spin gap and 
the frequency of the (in)commensurate peaks to decrease.
The quantitative role played by $U$ can be illustrated by reproducing the
frequency dependences of the RPA susceptibility in 
Fig.~%
\ref{Fig: Frequency leading edge of chi''RPA moving to left with underdoping}
with all BCS parameters unchanged but with $U$ fixed to its value at
optimal doping. The result is displayed in 
Fig.~\ref{Fig: U constant. No higher harmonics}a 
and
Fig.~\ref{Fig: U constant. No higher harmonics}b 
where one sees that: 
(i) The resonance energy moves with underdoping to higher 
energies as a result of the increasing gap maximum in panel (a),
and
(ii) the leading edge    moves with underdoping to lower
energies for the wave vector
$(1.26\pi,\pi)$ in panel~(b).
(The position of the leading edge is defined to be at the 
half maximum of the edge on the low frequency side.)
We see that the spin gap value
at a wave vector sufficiently far away from $(\pi,\pi)$
is rather insensitive to
keeping $U$ constant as in
Fig.~\ref{Fig: U constant. No higher harmonics}b
or to adjusting $U$ so as to get the correct resonance energy as in 
Fig.~%
\ref{Fig: Frequency leading edge of chi''RPA moving to left with underdoping}b.
We conclude that the dominant doping dependence of the 
\mbox{$\mathbf{q}$-dependent} spin gap 
comes from the doping dependence
of the higher harmonics for a wave vector sufficiently far away from
$(\pi,\pi)$.

This insensitivity
depends in a crucial way on how large $U$ becomes with underdoping.
This can be illustrated in a rather dramatic way by
switching off the BCS parameter $\Delta^{\ }_2$ for all dopings.
The parameter $\Delta^{\ }_1$ is then
chosen so that the maximum of the superconducting gap
on the Fermi surface $\Delta^{\ }_0$ agrees with the
values in Ref.~\onlinecite{mesot99},
i.e., we find that  $\Delta^{\ }_1$ equals 
$18.8meV$, 
$19.9 meV$, 
$20.5 meV$, and 
$22.1meV$, 
for samples with a $T_c$ of $87K$, $83K$, $75K$, and $68K$, respectively.
For the coupling constant $U$ we demand that its value is chosen so as 
to reproduce the same
resonances as in 
Fig.~%
\ref{Fig: Frequency leading edge of chi''RPA moving to left with underdoping}a,
i.e., we find that $U$ takes the values
$170meV$,
$184meV$,
$196meV$,
and
$211meV$,
when $T^{\ }_{c}$ takes the values
$87K$,
$83K$,
$75K$,
and
$68K$,
respectively.
Evidently, the increase of $U$ with underdoping is now much stronger
than in 
table~\ref{table: BCS dispersion parameters from ARPES}.
Having recalibrated $U$ to the new parameters of the BCS
dispersion, we plot in 
Fig.~\ref{Fig: U constant. No higher harmonics}c
the frequency dependence
of $\chi^{\prime\prime}(\omega+i\Gamma,\mathbf{q})$
for the wave vector $(1.26\pi, \pi)$.
Comparison of 
Fig.~\ref{Fig: U constant. No higher harmonics}c
with
Fig.~\ref{Fig: Frequency leading edge of chi''RPA moving to left with underdoping}b
shows that a broad peak at optimal doping can be turned into
a resonance with underdoping due to a too strong increase in $U$.

To summarize, we expect that the $\mathbf{q}$-dependent spin gap
always decreases with underdoping. 
Far away from the antiferromagnetic wave vector $(\pi, \pi)$ 
this is mostly a consequence of the decreasing slope of 
the gap function at the node, whereas close to $(\pi, \pi)$, 
it is a result of both the increasing coupling constant $U$ and the increasing
higher harmonics.
This is confirmed by the numerical simulations presented in 
Sec.~\ref{sec: Numerical results}. The effect of a decreasing
$\mathbf{q}$-dependent spin gap upon 
changing the gap parameters induced by underdoping is present in all 
fermiology scenarios. For example,
we have also performed calculations with a residual
nearest-neighbor antiferromagnetic interaction
instead of the on-site Hubbard repulsion
as well as within a RPA treatment of the bare propagator
of collective spin-1 excitations weakly coupled to BCS
quasiparticles in the spirit of 
Refs.\ \onlinecite{morr98,abanov99,morr00,pines00,chubukov01}.
In both cases we observe a softening of the 
$\mathbf{q}$-dependent spin gap 
as a result of switching on higher-order 
\mbox{$d$-wave} gap harmonics which is
rooted in the same mechanism as for the single-band Hubbard model.

\section{Conclusions}
\label{sec: Summary}

In conclusion, we have examined the effects of higher
\mbox{$d$-wave} 
gap harmonics induced by underdoping on the dynamical magnetic 
susceptibility of 
high-$T_c$ cuprates based on a fermiology approach. 
The calculations 
are carried out for a single-band Hubbard model with an on-site repulsion 
treated within the random phase approximation.
The input parameters for the BCS dispersions are taken directly from angle 
resolved photoemission measurements on Bi2212. We find that the inclusion of
higher harmonics decreases the  $\mathbf{q}$-dependent spin gap 
to a degree consistent with
experiments performed on YBCO. This effect is robust in that it does not depend
on the detailed nature of the fermiology model.
The downward dispersion of the incommensurate peaks 
is reproduced and shown to move down 
in energy with underdoping. We find a crossover from parallel to diagonal 
incommensuration above the resonance energy. However, this effect depends 
sensitively on the shape of the BCS dispersions and on the details of the 
fermiology model, here on the assumption of an on-site Hubbard residual 
interaction. With the advent of large enough
Bi2212 samples for neutron studies it will be possible to compare these 
predictions to experiments.

\section*{ACKNOWLEDGMENTS}
We would like to thank Hiroyuki Yamase and Pengcheng Dai for 
valuable discussions. This work was supported in parts by the
Swiss National Science Foundation under grant No.\ 200021-101765/1.

\appendix

\section{Degeneracies of  
$\mathrm{min}^{\ }_{\mathbf{k}}\,E^{\ }_2(\mathbf{q},\mathbf{k})$
         }
\label{appsec: Degeneracy}

\unitlength1cm
 \begin{figure}[t] 
 \begin{picture}(0,0)
\put(-4,-.7){\bf a)}
\put(.1,-.7){\bf b)}
\put(-4,-4.6){\bf c)}
\put(.1,-4.6){\bf d)}
\end{picture}
\begin{center}
\includegraphics[width=0.22\textwidth]{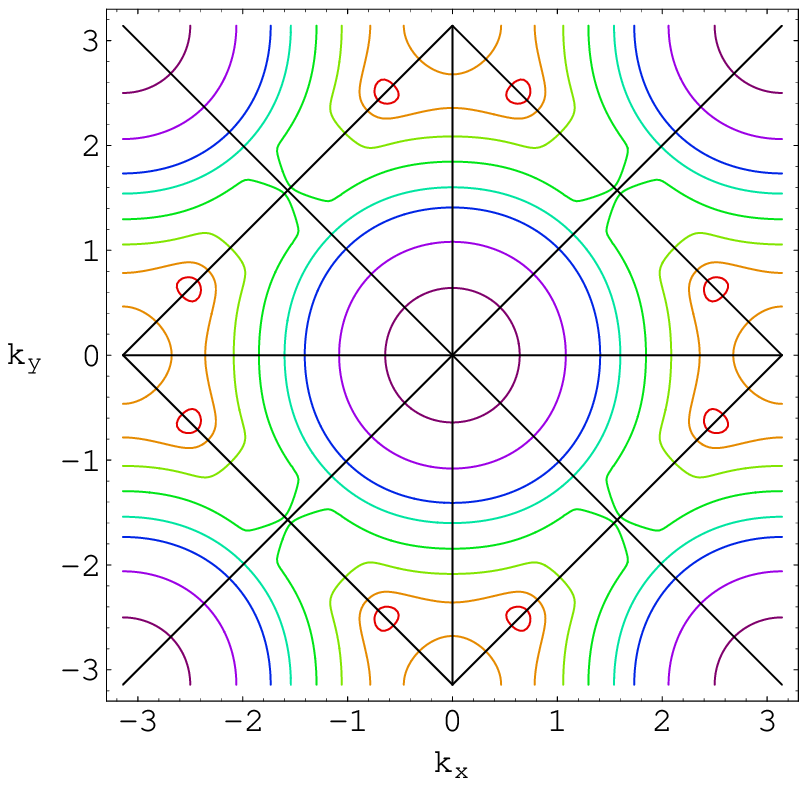}
\includegraphics[width=0.22\textwidth]{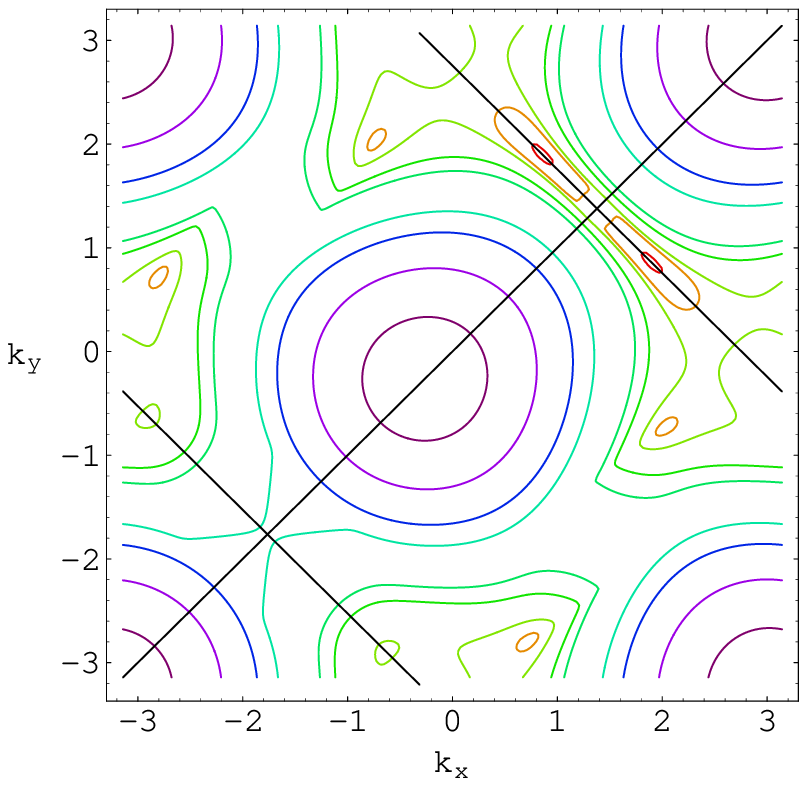}
\includegraphics[width=0.22\textwidth]{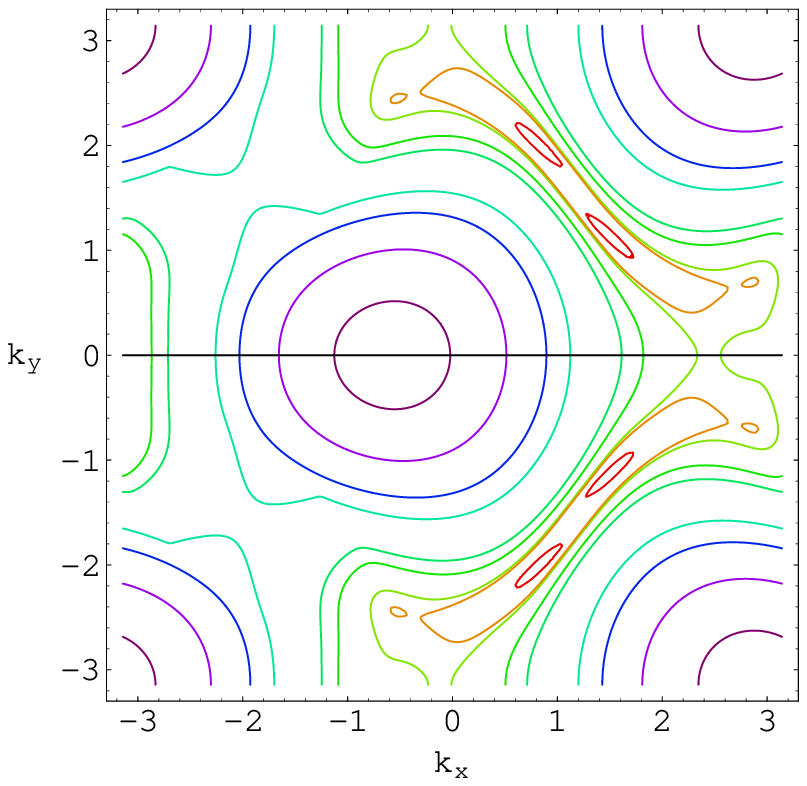}
\includegraphics[width=0.22\textwidth]{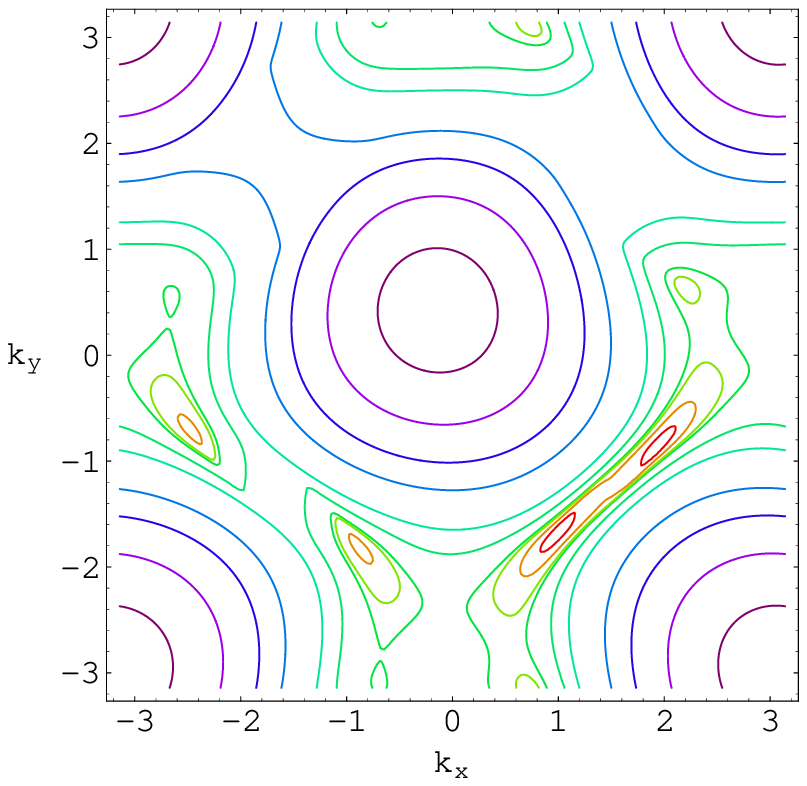}
\caption{
Panel (a) displays the  
$\mathbf{k}$~dependence of~$E^{\ }_2(\mathbf{q}, \mathbf{k})$
at the~$\mathbf{q}$-vector~$(\pi, \pi)$ in the first Brillouin 
zone~$\left] -\pi, \pi \right] \times \left] -\pi, \pi \right]$
for the BCS dispersion of Bi2212 at optimal doping 
(see table \ref{table: BCS dispersion parameters from ARPES}).
Panels (b), (c), and (d) are the same as panel (a) except for the momentum
transfer~$\mathbf{q}=(1.1225\pi, 1.1225\pi)$, 
         $\mathbf{q}=(1.26\pi, \pi)$, 
     and $\mathbf{q}=(1.065\pi, 0.805\pi)$, respectively.
The elevation of the contour lines increases with color, 
red being the lowest and violet the highest. 
The axes of reflection are drawn in black.
There are one \mbox{8-fold} degenerate minimum in panel (a), 
two \mbox{2-fold} degenerate minima and one 
\mbox{4-fold} degenerate minimum in panel (b), 
two \mbox{4-fold} degenerate minima in panel (c), 
and four \mbox{2-fold} degenerate minima in panel (d).
        }
\label{fig: BZ}
\end{center}
\end{figure}

The 2-particle dispersion $E^{\ }_2(\mathbf{q},\mathbf{k})$ is 
$2\pi$-periodic in all four variables.
In the following, when we refer to the first Brillouin zone, 
we have in mind 
 the \mbox{$\mathbf{k}$-vectors} that belong to the 
 primitive cell~$\left] -\pi, \pi \right] \times \left] -\pi, \pi \right]$.
For any wave vector~$\mathbf{q}$, $E^{\ }_2(\mathbf{q},\mathbf{k})$ 
is invariant under the involutive 
symmetry transformation
\begin{subequations}
\begin{eqnarray} \label{eq: symmetry trafo}
(k_x, k_y) \to (-k_x -q_x, -k_y -q_y) ,
\end{eqnarray}
which can be viewed as a rotation of angle~$\pi$ about the point 
\begin{eqnarray} \label{eq: origin of rotation}
(k_x, k_y) =-\left(\frac{q_x}{2}, \frac{q_y}{2}\right) .
\end{eqnarray}
\end{subequations}
For special values of $\mathbf{q}$ the \mbox{$2$-particle} 
dispersion possesses additional symmetries.

At the wave vector $\mathbf{q}= (\pi, \pi)$, $E_2$ 
is left invariant by the point-group transformations of the  
two-dimensional square lattice. 
Together with transformation~(\ref{eq: symmetry trafo}) the 
eight elements of the point group form a group of 16 elements. 
Hence, the two-dimensional
primitive cell in $\mathbf{k}$ space can be divided into 16 
triangular subcells which
transform into each other under the action of the symmetry group. 
As the \mbox{$2$-particle}
energy is a continuous and bounded function in~$\mathbf{k}$, 
$E_2(\pi, \pi, \mathbf{k})$
must have at least one minimum and one maximum in each subcell. Furthermore,
since the sides of the subcells are mirror axes, the gradient of the 
\mbox{$2$-particle} dispersion 
at $\mathbf{q}=(\pi, \pi)$ vanishes on every corner of the triangles.  
The degeneracy of the (local) minima of~$E_2(\pi, \pi, \mathbf{k})$ in 
the first Brillouin
zone is a direct consequence of the symmetry transformations. 
Depending on whether 
the minima lie in the interior of the triangles, on the reflection axes, 
on corners which are 
shared by four triangles or on corners which are shared by eight triangles,
 the (local) minima of the \mbox{2-particle} energy are either 16-fold, 
8-fold, 
4-fold, 
or 2-fold degenerate, respectively. 
For the BCS dispersion of Bi2212 at optimal
doping 
(see table~\ref{table: BCS dispersion parameters from ARPES}) 
the symmetries and the degeneracy of the global minimum 
of~$E_2(\pi, \pi, \mathbf{k})$ are depicted in Fig.~\ref{fig: BZ}a. 
  
When $\mathbf{q}$ is chosen to be on either of the two diagonals passing 
through~$(\pi, \pi)$, the \mbox{2-particle} 
energy is symmetric under the reflection 
about the corresponding diagonal in the Brillouin zone.
Together with transformation~(\ref{eq: symmetry trafo}) 
this yields a symmetry group with four 
elements. The first Brillouin zone decays into four polygonal subcells 
which are related to each 
other by the action of the symmetry group. Hence, 
the degeneracy of the (local) minima is 4-fold, 
when they are in the interior of the polygons, 2-fold, 
when they lie on a reflection axis and 1-fold,
when they are on a corner of the subcells. 
This is illustrated in Fig.~\ref{fig: BZ}b for the case 
when~$\mathbf{q}$ is on the diagonal 
connecting~$(-\pi, \pi)$ and~$(\pi, \pi)$. 

If $\mathbf{q}$ is located on the horizontal (vertical) 
axis passing through the commensurate 
point $(\pi, \pi)$, $E_2$ 
is left invariant under the reflection about the horizontal (vertical) axis 
which goes through the origin of the Brillouin zone. 
The (local) minima are doubly degenerate, 
when they lie on the reflection axis or on the 
point~(\ref{eq: origin of rotation}); 
otherwise they are 4-fold degenerate (see Fig.~\ref{fig: BZ}c).

Finally, when $\mathbf{q}$ is chosen not to be on any of the above symmetry 
axes, the symmetries of the \mbox{2-particle} energy 
are reduced to rotation~(\ref{eq: symmetry trafo}). 
The degeneracy of the (local) minima is 1-fold, when they lie on the symmetry 
point~(\ref{eq: origin of rotation}), 
and 2-fold otherwise (see Fig.~\ref{fig: BZ}d).

From the continuity of $E_2(\mathbf{q}, \mathbf{k})$, now considered
as a function of $\mathbf{q}$,
it follows that the number of local minima of
the \mbox{2-particle} energy in the first Brillouin zone for 
the~$\mathbf{k}$'s 
is constant on an open neighborhood
of~$\mathbf{q}=(\pi, \pi)$. For the BCS dispersions of 
table~\ref{table: BCS dispersion parameters from ARPES}
this neighborhood includes the area around the antiferromagnetic
wave vector where most of the spectral weight of
$\chi^{\prime \prime}_{\mathrm{RPA}}(\omega, \mathbf{q})$ is located,
as is illustrated in Fig.\ \ref{fig: BZ}.  
With these hopping and gap parameters the minima of the \mbox{2-particle} 
energy at the commensurate point~$(\pi, \pi)$
are 8-fold degenerate, since  they lie on the magnetic 
Brillouin zone boundary. 
There is one doubly and one quadruple degenerate local minimum, 
and a doubly degenerate global 
minimum,  when the momentum transfer lies on a diagonal. 
If~$\mathbf{q}$ is chosen to be on a 
vertical or horizontal axis passing through the 
commensurate point~$(\pi, \pi)$, the two (local) minima 
are 4-fold degenerate.  Finally, when the wave vector~$\mathbf{q}$ 
is not on any of the symmetry 
axes, there are four doubly degenerate (local) minima.


\begin{thebibliography}{999}





\bibitem{rossat91a}
J.\ Rossat-Mignod, L.\ P.\ Regnault, C.\ Vettier, P.\ Bourges, P.\ Burlet, 
J.\ Bossy, J.\ Y.\ Henry, and G.\ Lapertot,
Physica C \textbf{185}, 86 (1991).

\bibitem{mook93}
H.\ A.\ Mook, M.\ Yethiraj, G.\ Aeppli, T.\ E.\ Mason, and T.\ Armstrong,
Phys.\ Rev.\ Lett.\ \textbf{70}, 3490 (1993).

\bibitem{sternlieb94}
B.\ J.\ Sternlieb, J.\ M.\ Tranquada, G.\ Shirane, M.\ Sato, and S.\ Shamoto,
Phys.\ Rev.\ B \textbf{50}, 12915 (1994).

\bibitem{fong95}
H.\ F.\ Fong, B.\ Keimer, P.\ W.\ Anderson, D.\ Reznik, F.\ Dogan, 
and I.\ A.\ Aksay,
Phys.\ Rev.\ Lett.\ \textbf{75}, 316 (1995).

\bibitem{dai96}
P.\ Dai, M.\ Yethiraj, H.\ A.\ Mook, T.\ B.\ Lindemer, and F.\ Dogan,
Phys.\ Rev.\ Lett.\ \textbf{77}, 5425 (1996).

\bibitem{bourges96}
P.\ Bourges, L.\ P.\ Regnault, Y.\ Sidis, and C.\ Vettier,
Phys.\ Rev.\ B \textbf{53}, 876 (1996).

\bibitem{fong97}
H.\ F.\ Fong, B.\ Keimer, D.\ L.\ Milius, and I.\ A.\ Aksay,
Phys.\ Rev.\ Lett.\ {\bf 78}, 713 (1997).

\bibitem{dai98}
P.\ Dai , H.\ A.\ Mook, and F.\ Dogan,
Phys.\ Rev.\ Lett.\ \textbf{80}, 1738 (1998).

\bibitem{mook98}
H.\ A.Mook, P.\ Dai, S.\ M.\ Hayden, G.\ Aeppli, T.\ G.\ Perring, 
and F.\ Dogan,
Nature \textbf{395}, 580 (1998).

\bibitem{dai99}
P.\ Dai, H.\ A.\ Mook, S.\ M.\ Hayden, G.\ Aeppli, T.\ G.\ Perring, 
R.\ D.\ Hunt, and F.\ Dogan,
Science \textbf{284}, 1344 (1999).

\bibitem{arai99}  
M.\ Arai, T.\ Nishijima, Y.\ Endoh, T.\ Egami, S.\ Tajima, K.\ Tomimoto, 
Y.\ Shiohara, M.\ Takahashi, A.\ Garrett, and S.\ M.\ Bennington,
Phys.\ Rev.\ Lett.\ \textbf{83}, 608 (1999).

\bibitem{bourges00}  
P.\ Bourges, Y.\ Sidis, H.\ F.\ Fong, L.\ P.\ Regnault, J.\ Bossy, 
A.\ Ivanov, and B.\ Keimer,
Science \textbf{288}, 1234 (2000).

\bibitem{fong00}  
H.\ F.\ Fong, P.\ Bourges, Y.\ Sidis, L.\ P.\ Regnault, J.\ Bossy, 
A.\ Ivanov, D.\ L.\ Milius, I.\ A.\ Aksay, and B.\ Keimer,
Phys.\ Rev.\ B \textbf{61}, 14773 (2000).

\bibitem{mook00} 
H.\ A.\ Mook, P.\ Dai, F.\ Do{\u g}an, and R.\ D.\ Hunt,
Nature \textbf{404}, 729 (2000).

\bibitem{dai01}  
P.\ Dai, H.\ A.\ Mook, R.\ D.\ Hunt, and F.\ Do{\u g}an,
Phys.\ Rev.\ B \textbf{63}, 054525 (2001).

\bibitem{pailhes03} 
S.\ Pailhes, Y.\ Sidis, P.\ Bourges, C.\ Ulrich, V.\ Hinkov, L.\ P.\ 
Regnault, A.\ Ivanov, B.\ Liang, C.\ T.\ Lin, C.\ Bernhard, and B.\ Keimer,
Phys.\ Rev.\ Lett.\ \textbf{91}, 237002 (2003).

\bibitem{stock04} 
C.\ Stock, W.\ J.\ L.\ Buyers, R.\ Liang, D.\ Peets, Z.\ Tun, D.\ Bonn, 
W.\ N.\ Hardy, and R.\ J.\ Birgeneau,
Phys.\ Rev.\ B  \textbf{69}, 014502 (2004).

\bibitem{pailhes04}
S.\ Pailhes, Y.\ Sidis, P.\ Bourges, V.\ Hinkov, A.\ Ivanov, C.\ Ulrich, 
L.\ P.\ Regnault, and B.\ Keimer,
preprint cond-mat/0403609.

\bibitem{hayden04}
S.~M.~Hayden, H.~A.~Mook, P.~Dai, T.~G.~Perring, and F.~Dogan,
Nature (in press).


\bibitem{shirane89}
G.\ Shirane, R.\ J.\ Birgeneau, Y.\ Endoh, P.\ Gehring,
M.\ A.\ Kastner, K.\ Kitazawa, H.\ Kojima, I.\ Tanaka,
T.\ R.\ Thurston, and K.\ Yamada,
Phys.\ Rev.\ Lett.\ \textbf{63}, 330 (1989).

\bibitem{cheong91}
S.\ W.\ Cheong, G.\ Aeppli, T.\ E.\ Mason, H.\ Mook, S.\ M.\ Hayden, 
P.\ C.\ Canfield, Z.\ Fisk, K.\ N.\ Clausen, and J.\ L.\ Martinez,
Phys.\ Rev.\ Lett.\ \textbf{67}, 1791 (1991).

\bibitem{mason93}
T.\ E.\ Mason, G.\ Aeppli, S.\ M.\ Hayden, A.\ P.\ Ramirez, and H.\ A.\ Mook,
Phys.\ Rev.\ Lett.\ \textbf{71}, 919 (1993).

\bibitem{yamada95}
K.\ Yamada,  S.\ Wakimoto, G.\ Shirane, C.\ H.\ Lee, M.\ A.\ Kastner, 
S.\ Hosoya, M.\ Greven, Y.\ Endoh, and R.\ J.\ Birgeneau,
Phys.\ Rev.\ Lett.\ \textbf{75}, 1626 (1995).

\bibitem{hayden96} 
S.\ M.\ Hayden, G.\ Aeppli, H.\ A.\ Mook, T.\ G.\ Perring, T.\ E.\ Mason, 
S.-W.\ Cheong, and Z.\ Fisk, 
Phys.\ Rev.\ Lett. \textbf{76}, 1344 (1996).

\bibitem{mason96}
T.\ E.\ Mason, A.\ Schr\"oder, G.\ Aeppli, H.\ A.\ Mook, and S.\ M.\ Hayden,
Phys.\ Rev.\ Lett. \textbf{77}, 1604 (1996).

\bibitem{aeppli97}
G.\ Aeppli, T.\ E.\ Mason, S.\ M.\ Hayden, H.\ A.\ Mook, and J.\ Kulda,
Science \textbf{278}, 1432 (1997).

\bibitem{yamada98}
K.\ Yamada, C.\ H.\ Lee, K.\ Kurahashi, J.\ Wada, S.\ Wakimoto, S.\ Ueki, 
H.\ Kimura, Y.\ Endoh, S.\ Hosoya, G.\ Shirane, R.\ J.\ Birgeneau, 
M.\ Greven, M.\ A.\ Kastner, and Y.\ J.\ Kim,
Phys.\ Rev.\ B \textbf{57}, 6165 (1998).

\bibitem{lake99}
B.\ Lake, G.\ Aeppli, T.\ E.\ Mason, A.\ Schr\"oder, D.\ F.\ McMorrow, 
K.\ Lefmann, M.\ Isshiki, M.\ Nohara, H.\ Takagi, and S.\ M.\ Hayden,
Nature \textbf{400}, 43 (1999).

\bibitem{lee00}
C.\ H.\ Lee, K.\ Yamada, Y.\ Endoh, G.\ Shirane, R.\ J.\ Birgeneau, 
M.\ A.\ Kastner, M.\ Greven, and Y.\ J.\ Kim,
J.\ Phys.\ Soc.\ Jpn.\ \textbf{69}, 1170 (2000).

\bibitem{gilardi04}
R.\ Gilardi, S.\ Streule, A.\ Hiess, H.\ M.\ Ronnow, M.\ Oda, N.\ Momono, 
M.\ Ido, and J.\ Mesot, 
in print in Physica B.

\bibitem{christensen04}
N.\ B.\ Christensen, D.\ F.\ McMorrow, H.\ M.\ R\o nnow, B.\ Lake, 
S.\ M.\ Hayden, G.\ Aeppli, T.\ G.\ Perring, M.\ Mangkorntong, M.\ Nohara, 
and H.\ Tagaki,
preprint cond-mat/0403439.

\bibitem{fong99}
H.\ F.\ Fong, P.\ Bourges, Y.\ Sidis, L.\ P.\ Regnault, A.\ Ivanov, 
G.\ D.\ Gul, N.\ Koshizuka, and B.\ Keimer,
Nature \textbf{398}, 588 (1999).
\bibitem{mook98bis}
H.\ A.\ Mook, F.\ Dogan, and B.\ C.\ Chakoumakos,
unpublished, cond-mat/9811100.
\bibitem{mesot00}
J.\ Mesot, N.\ Metoki, M.\ B\"ohm, A.\ Hiess, and K.\ Kadowaki,
Physica C \textbf{341}, 2105 (2000).
\bibitem{he01}
H.\ He,  Y.\ Sidis, P.\ Bourges, G.\ D.\ Gu, A.\ Ivanov, N.\ Koshizuka, 
B.\ Liang, C.\ T.\ Lin, L.\ P.\ Regnault, E.\ Schoenherr, and B.\ Keimer,
Phys.\ Rev.\ Lett.\ \textbf{86}, 1610 (2001).


\bibitem{he02}
H.\ He, P.\ Bourges, Y.\ Sidis, C.\ Ulrich, L.\ P.\ Regnault, S.\ Pailhes, 
N.\ S.\ Berzigiarova, N.\ N.\ Kolesnikov, and B.\ Keimer,
Science \textbf{295}, 1045 (2002).

\bibitem{batista01}
C.\ D.\ Batista, G.\ Ortiz, and A.\ V.\ Balatsky,
Phys.\ Rev.\ B \textbf{64}, 172508 (2001).

\bibitem{tranquada04}
J.\ M.\ Tranquada, H.\ Woo, T.\ G.\ Perring, H.\ Goka, G.\ D.\ Gu, G.\ Xu, 
M.\ Fujita, and K.\ Yamada,
preprint cond-mat/0401621.



\bibitem{machida89}
K.\ Machida,
Physica C \textbf{158}, 192 (1989).

\bibitem{poilblanc89}
D.\ Poilblanc and T.\ M.\ Rice,
Phys.\ Rev.\ B \textbf{39}, R9749 (1989).

\bibitem{zaanen89}
J.\ Zaanen and O.\ Gunnarsson,
Phys.\ Rev.\ B \textbf{40}, R7391 (1989).

\bibitem{schulz89}
H.\ J.\ Schulz,
J.\ Phys.\ (Paris) \textbf{50}, 2833 (1989).

\bibitem{emery93}
V.\ J.\ Emery and S.\ A.\ Kivelson,
Physica C \textbf{209}, 597 (1993).

\bibitem{kivelson03}
For a review, see
S.\ A.\ Kivelson, I.\ P.\ Bindloss, E.\ Fradkin, V.\ Oganesyan, J.\ M.\ 
Tranquada, A.\ Kapitulnik, and C.\ Howald,
Rev.\ Mod.\ Phys.\  \textbf{75}, 1201 (2003).

\bibitem{uhrig04}
G.\ S.\ Uhrig, K.\ P.\ Schmidt, M.\ Gr\"uninger,
preprint cond-mat/0402659


\bibitem{demler95}
E.\ Demler and S.\ C.\ Zhang,
Phys.\ Rev.\ Lett.\ \textbf{75}, 4126 (1995).

\bibitem{zhang97}
S.\ C.\ Zhang,
Science \textbf{275}, 1089 (1997).

\bibitem{greiter97}
Martin Greiter,
Phys.\ Rev.\ Lett.\ \textbf{79}, 4898 (1997).

\bibitem{demler98}
E.\ Demler, H.\ Kohno, and S.-C.\ Zhang,
Phys.\ Rev.\ B \textbf{58}, 5719 (1998).

\bibitem{brinckmann98}
J.\ Brinckmann and P.\ A.\ Lee,
J.\ Phys.\ Chem.\ Solids \textbf{59}, 1811 (1998).

\bibitem{tschernyshyov01}
O.\ Tchernyshyov, M.\ R.\ Norman, and A.\ V.\ Chubukov,
Phys.\ Rev.\ B \textbf{63}, 144507 (2001).

\bibitem{jiang-ping01}
Jiang-Ping Hu and Shou-Cheng Zhang,
  Phys.\ Rev.\ B \textbf{64}, 100502(R) (2001).


\bibitem{bulut90}
N.\ Bulut, D.\ W.\ Hone, D.\ J.\ Scalapino, and N.\ E.\ Bickers,
Phys.\ Rev.\ B \textbf{41}, 1797 (1990).

\bibitem{schulz90}
H.\ J.\ Schulz,
Phys.\ Rev.\ Lett.\ \textbf{64}, 1445 (1990).

\bibitem{bulut92}
N.\ Bulut and D.\ J.\ Scalapino,
Phys.\ Rev.\ Lett.\ \textbf{68}, 706 (1992).

\bibitem{bulut93}
N.\ Bulut and D.\ J.\ Scalapino,
Phys.\ Rev.\ B \textbf{47}, 3419 (1993).

\bibitem{benard93}
P.\ B\'enard, L.\ Chen, and A.-M.\ S.\ Tremblay,
Phys.\ Rev.\ B \textbf{47}, 15217 (1993).

\bibitem{lavagna94}
M.\ Lavagna and G.\ Stemmann,
Phys.\ Rev.\ B \textbf{49}, 4235 (1994).

\bibitem{marel95}
D.\ van der Marel,
Phys.\ Rev.\ B \textbf{51}, 1147 (1995).

\bibitem{bulut96}
N.\ Bulut and D.\ J.\ Scalapino,
Phys.\ Rev.\ B \textbf{53}, 5149 (1996).

\bibitem{salkola98}
M.\ I.\ Salkola and J.\ R.\ Schrieffer,
Phys.\ Rev.\ B \textbf{58}, R5944 (1998).

\bibitem{norman00}
M.\ R.\ Norman,
Phys.\ Rev.\ B \textbf{61}, 14751 (2000).

\bibitem{norman01}
M.\ R.\ Norman,
Phys.\ Rev.\ B \textbf{63}, 092509 (2001).


\bibitem{maki94}
K.\ Maki and H.\ Won,
Phys.\ Rev.\ Lett.\ \textbf{72}, 1758 (1994).

\bibitem{mazin95}
I.\ I.\ Mazin and V.\ M.\ Yakovenko,
Phys.\ Rev.\ Lett.\ \textbf{75}, 4134 (1995).

\bibitem{blumberg95}
G.\ Blumberg, B.\ P.\ Stojkovic, and M.\ V.\ Klein,
Phys.\ Rev.\ B \textbf{52}, R15741 (1995).

\bibitem{li98}
J.-X.\ Li, W.-G.\ Yin, and C.-D.\ Gong,
Phys.\ Rev.\ B \textbf{58}, 2895 (1998).


\bibitem{tanamoto91}
T.\ Tanamoto, K.\ Kuboki, and H.\ Fukuyama,
J.\ Phys.\ Soc.\ Jpn.\ \textbf{60}, 3072 (1991).

\bibitem{tanamoto93}
T.\ Tanamoto, H.\ Kohno, and H.\ Fukuyama,
J.\ Phys.\ Soc.\ Jpn.\ \textbf{62}, 717 (1993).

\bibitem{tanamoto94}
T.\ Tanamoto, H.\ Kohno, and H.\ Fukuyama,
J.\ Phys.\ Soc.\ Jpn.\ \textbf{63}, 2739 (1994).

\bibitem{stemmann94}
G.\ Stemmann, C.\ P{\'e}pin, and M.\ Lavagna,
Phys.\ Rev.\ B \textbf{50}, 4075 (1994).

\bibitem{brinckmann99}
J.\ Brinckmann and P.\ A.\ Lee,
Phys.\ Rev.\ Lett.\ \textbf{82}, 2915 (1999).

\bibitem{yamase00}
H.\ Yamase and H.\ Kohno,
J.\ Phys.\ Soc.\ Jpn.\ \textbf{69}, 2151 (2000).

\bibitem{li00}
J.-X.\ Li, C.-Y.\  Mou, and T.\ K.\ Lee,
Phys.\ Rev.\ B \textbf{62}, 640 (2000).

\bibitem{yamase01}
H.\ Yamase and H.\ Kohno,
J.\ Phys.\ Soc.\ Jpn.\ \textbf{70}, 2733 (2001).

\bibitem{brinckmann01}
J.\ Brinckmann and P.\ A.\ Lee,
Phys.\ Rev.\ B \textbf{65}, 014502 (2001).

\bibitem{yamase02}
H.\ Yamase,
J.\ Phys.\ Soc.\ Jpn.\ \textbf{71}, 1154 (2002).

\bibitem{li02}
J.-X.\ Li, and C.-D.\ Gong,
Phys.\ Rev.\ B \textbf{66}, 014506 (2002).

\bibitem{yamase03}
H.\ Yamase and H.\ Kohno,
Phys.\ Rev.\ B \textbf{68}, 014502 (2003).

\bibitem{li03}
J.-X.\ Li, J.\ Zhang, and J.\ Luo,
Phys.\ Rev.\ B \textbf{68}, 224503 (2003).


\bibitem{onufrieva95}
F.\ Onufrieva and J.\ Rossat-Mignod,
Phys.\ Rev.\ B \textbf{52}, 7572 (1995).

\bibitem{onufrieva99}
F.\ Onufrieva, P.\ Pfeuty, and M.\ Kiselev,
Phys.\ Rev.\ Lett.\ \textbf{82}, 2370 (1999).

\bibitem{onufrieva00}
F.\ Onufrieva and P.\ Pfeuty,
Phys.\ Rev.\ B \textbf{61}, 799 (2000).

\bibitem{onufrieva02}
F.\ Onufrieva and P.\ Pfeuty,
Phys.\ Rev.\ B \textbf{65}, 054515 (2002).


\bibitem{morr98}
D.\ K.\ Morr and D.\ Pines,
Phys.\ Rev.\ Lett.\ \textbf{81}, 1086 (1998).

\bibitem{abanov99}
A.\ Abanov and A.\ V.\ Chubukov,
Phys.\ Rev.\ Lett.\ \textbf{83}, 1652 (1999).

\bibitem{morr00}
D.\ K.\ Morr and D.\ Pines,
Phys.\ Rev.\ B.\ \textbf{61}, R6483 (2000).

\bibitem{pines00}
D.\ K.\ Morr and D.\ Pines,
Phys.\ Rev.\ B.\ \textbf{62}, 15177 (2000).

\bibitem{chubukov01}
A.\ V.\ Chubukov, B.\ Janko, and O.\ Tchernyshyov,
Phys.\ Rev.\ B.\ \textbf{63}, R180507 (2001).


\bibitem{si93}
Q.\ Si, Y.\ Zha, K.\ Levin, and J.\ P.\ Lu,
Phys.\ Rev.\ B \textbf{47}, 9055 (1993).

\bibitem{zha93}
Y.\ Zha, K.\ Levin, and Q.\ Si,
Phys.\ Rev.\ B \textbf{47}, 9124 (1993).

\bibitem{liu95}
D.\ Z.\ Liu, Y.\ Zha, and K.\ Levin,
Phys.\ Rev.\ Lett.\ \textbf{75}, 4130 (1995).

\bibitem{millis96}
A.\ J.\ Millis and H.\ Monien,
Phys.\ Rev.\ B \textbf{54}, 16172 (1996).

\bibitem{kao00}
Y.-J.\ Kao, Q.\ Si, and K.\ Levin,
Phys.\ Rev.\ B \textbf{61}, R11898 (2000).


\bibitem{pao95}
C.-H.\ Pao and N.\ E.\ Bickers,
Phys.\ Rev.\ B \textbf{51}, 16310 (1995).

\bibitem{dahm96}
T.\ Dahm, D.\ Manske, and L.\ Tewordt,
Phys.\ Rev.\ B \textbf{54}, 6640 (1996).

\bibitem{takimoto98}
T.\ Takimoto and T.\ Moriya,
J.\ Phys.\ Soc.\ Jpn.\ \textbf{67}, 3570 (1998).

\bibitem{dahm98}
T.\ Dahm, D.\ Manske, and L.\ Tewordt,
Phys.\ Rev.\ B \textbf{58}, 12454 (1998).

\bibitem{kuroki99}
K.\ Kuroki, R.\ Arita, and H.\ Aoki,
Phys.\ Rev.\ B \textbf{60}, 9850 (1999).

\bibitem{manske01}
D.\ Manske, I.\ Eremin, and K.\ H.\ Bennemann,
Phys.\ Rev.\ B \textbf{63}, 054517 (2001).


\bibitem{littlewood93}
P.\ B.\ Littlewood, J.\ Zaanen, G.\ Aeppli, and H.\ Monien,
Phys.\ Rev.\ B \textbf{48}, 487 (1993).

\bibitem{ito04}
M.\ Ito, Y.\ Yasui, S.\ Iikubo, M.\ Soda,
A.\ Kobayashi, M.\ Sato, K.\ Kakurai, C.-H.\ Lee,
and  K.\ Yamada,
preprint cond-mat/0311058.


\bibitem{lu92}
J.\ P.\ Lu,
Phys.\ Rev.\ Lett.\ \textbf{68}, 125 (1992).

\bibitem{kee99}
H.-Y.\ Kee and Y.\ B.\ Kim,
Phys.\ Rev.\ B \textbf{59}, 4470 (1999).

\bibitem{voo00}
K.-K.\ Voo, W.\ C.\ Wu, J.-X.\ Li, and T.\ K.\ Lee,
Phys.\ Rev.\ B \textbf{61}, 9095 (2000).



\bibitem{harris96}
J.M.~Harris, Z.-X.\ Shen, P.\ J.\ White, D.\ S.\ Marshall, M.\ C.\ Schabel, 
J.\ N.\ Eckstein, and I.\ Bozovic,
Phys.\ Rev.\ B\ \textbf{54}, R15665 (1996).

\bibitem{mesot99}
J.\ Mesot, M.\ R.\ Norman, H.\ Ding, M.\ Randeria, J.\ C.\ Campuzano, 
A.\ Paramekanti, H.\ M.\ Fretwell, A.\ Kaminski, T.\ Takeuchi, T.\ Yokoya, 
T.\ Sato, T.\ Takahashi, T.\ Mochiku, and K.\ Kadowaki,
Phys.\ Rev.\ lett.\ \textbf{83}, 840 (1999).

\bibitem{borisenko02}
S.V.\ Borisenko, A.\ A.\ Kordyuk, T.\ K.\ Kim, S.\ Legner, K.\ A.\ Nenkov, 
M.\ Knupfer, M.\ S.\ Golden, J.\ Fink, H.\ Berger, and R.\ Follath,
Phys.\ Rev.\ B\ \textbf{66}, 140509(R) (2002).

\bibitem{higher harmonics not caused by disorder}
Here, we follow the interpretation from 
Ref.~\onlinecite{mesot99} that
the changes in the superconducting gap as measured by ARPES are 
intrinsic and not
caused by disorder effects.



\bibitem{footnote on bilayer issue}
The YBCO and Bi2212 families are stacked bilayers.
One expects that the quasiparticle dispersions close to the Fermi energy
split into bonding and antibonding bands. This was first seen in both
ARPES and neutron scattering on YBCO.\cite{schabel98,fong00,pailhes03}
However, in the underdoped and optimally doped regimes the prominent 
incommensurate and
commensurate peaks in the magnetic susceptibility are only observed
in the antibonding band. Peaks are observed in the bonding band but
  they
are very broad both as a function of energy and as a function of
momentum. Recently a resonant excitation in the even channel was 
found for overdoped
YBCO, albeit with a much smaller intensity than the odd channel 
resonance.\cite{pailhes03}
It seems most likely that the resonance in the bonding band is only present
in the overdoped regime and grows in intensity as one increases the 
doping concentration.
Within the fermiology approach it was shown in
Ref.~\onlinecite{brinckmann01}
(see also Refs.~\onlinecite{liu95,mazin95,blumberg95,bulut96,dahm96,millis96})
that an effective one-band model for the antibonding band is
sufficient to account for the incommensurate and commensurate peaks in
the YBCO and Bi2212 families, if one only considers optimally doped 
and underdoped
samples. We shall adopt the same point of view in this
paper.

\bibitem{schabel98}
M.~C.~Schabel, C.-H.\ Park, A.\ Matsuura, Z.-X.\ Shen, D.\ A.\ Bonn, 
X.\ Liang, and W.\ N.\ Hardy,
Phys.\ Rev.\ B\ \textbf{57}, 6090 (1998).


\bibitem{chubukov04}
A.\ V.\ Chubukov and M.\ R.\ Norman, preprint cond-mat/0402304.


\bibitem{norman95}
M.\ R.\ Norman, M.\ Randeria, H.\ Ding, and J.\ C.\ Campuzano,
Phys.\ Rev.\ B \textbf{52}, 615 (1995).
\bibitem{ding97}
H.~Ding, M.~R.~Norman, T.~Yokoya, T.~Takeuchi, M.~Randeria,                                                               
J.~C.~Campuzano, T.~Takahashi, T.~Mochiku, and                                                                      
K.~Kadowaki,
Phys.~Rev.~Lett. \textbf{78}, 2628 (1997).
\bibitem{kordyuk02}
A.\ A.\ Kordyuk, S.\ V.\ Borisenko, M.\ S.\ Golden, S.\ Legner, 
K.\ A.\ Nenkov, M.\ Knupfer, J.\ Fink, H.\ Berger, L.\ Forro, and R.\ Follath,
Phys.\ Rev.\ B \textbf{66}, 014502 (2002).

\bibitem{campuzano99}
J.\ C.\ Campuzano, H.\ Ding, M.\ R.\ Norman, H.\ M.\ Fretwell, M.\ 
Randeria, A.\ Kaminski, J.\ Mesot, T.\ Takeuchi, T.\ Sato, T.\ Yokoya, 
T.\ Takahashi, T.\ Mochiku, K.\ Kadowaki, P.\ Guptasarma, D.\ G.\ Hinks, 
Z.\ Konstantinovic, Z.\ Z.\ Li, and H.\ Raffy,
Phys.\ Rev.\ Lett.\ \textbf{83}, 3709 (1999).
\bibitem{zasadzinski02}
J.\ F.\ Zasadzinski, L.\ Ozyuzer, N.\ Miyakawa, K.\ E.\ Gray, 
D.\ G.\ Hinks, and C.\ Kendziora,
Phys.\ Rev.\ Lett.\ \textbf{87}, 067005 (2001).


\bibitem{footnote on definition of k_1 and k_2} 
The coordinate system $(k^{\ }_{1},k^{\ }_{2})$
is defined by demanding that 
$
0=
\partial^2 E^{\ }_2(\mathbf{q},\mathbf{k})
/
\partial k^{\ }_{1}\partial k^{\ }_{2}
$ 
when $\mathbf{k}=\mathbf{k}^{\ }_{n}$.



\end{thebibliography}
\end{document}